\documentclass[a4paper,11pt]{article}
\ifdefined\XeTeXrevision\else\pdfoutput=1\fi
\usepackage{jheppub}
\usepackage{mathrsfs}
\usepackage{booktabs}
\usepackage{orcidlink}
\numberwithin{equation}{section}
\hypersetup{pdftitle={Thermodynamics within and across holographic CFTs},
            pdfauthor={Limin Zeng},
            pdfsubject={Charged holographic thermodynamics in Gauss--Bonnet and Weyl-squared gravity}}

\title{Thermodynamics within and across holographic CFTs}
\author[a,b,c]{Limin Zeng\texorpdfstring{\orcidlink{0009-0003-5703-2687}}{}}
\affiliation[a]{School of Fundamental Physics and Mathematical Sciences,
 Hangzhou Institute for Advanced Study, University of Chinese Academy of Sciences, Hangzhou 310024, China}
\affiliation[b]{Institute of Theoretical Physics, Chinese Academy of Sciences,
 Beijing 100190, China}
\affiliation[c]{University of Chinese Academy of Sciences, Beijing 100049, China}
\emailAdd{zenglimin25@mails.ucas.ac.cn}

\abstract{Varying the bulk curvature coupling changes the boundary Weyl anomaly, so different couplings label distinct holographic conformal field theories (CFTs). We compare their charged black holes on a common boundary cylinder at fixed sphere radius, charge and proper bath temperature. In each CFT we minimize the canonical free energy over the horizon radius and subtract the common source times the coupling. Two theories coexist when their resulting potentials share the lowest value. The comparison is preserved under a change of energy prescription when the external control is transformed accordingly. In five bulk dimensions, concavity at fixed horizon radius selects the Gauss--Bonnet endpoints on the intervals studied. Concavity along thermal branches gives local endpoint selection near the Einstein transition when Weyl-squared is the sole correction to Einstein--Maxwell gravity. The entropy difference and coupling separation set the coexistence slope. A small--large transition in one CFT produces a triple point whose slope jump is fixed by the Wald entropy gap.}

\begin{document}
\maketitle
\section{Introduction}
\label{sec:intro}

At fixed coupling, the thermal saddles of one holographic conformal field theory (CFT) compete under a common anti-de Sitter (AdS) bulk action and boundary conditions \cite{Witten1998,Chamblin1999}. Along the families studied here, varying the bulk curvature coupling at fixed normalization changes the boundary CFT. This change is visible in its four-dimensional Weyl anomaly, whose Euler and Weyl-squared coefficients can vary independently \cite{HenningsonSkenderis1998,BuchelMyersSinha2009}.

We compare charged spherical saddles at a common boundary sphere radius, conserved charge and proper bath temperature in the unit-lapse cylinder clock. An external source contributes a term linear in the dimensionless curvature coupling. Since this term is constant within each theory, its role is to change the relative potentials of different theories. The boundary first law supplies the work term conjugate to the coupling \cite{Visser2022,Ahmed2023,Punia2024,CuiFan2024}. At each coupling we minimize the potential over the horizon radius. Two theories coexist where their minimized potentials are equal and lowest among the candidates. The resulting curve in the source--temperature plane obeys a Clapeyron relation. A small--large switch in one theory produces a kink at a triple point with three saddles from two CFTs. The jump in slope is set by the gap in Wald entropy, the horizon Noether-charge entropy in higher-curvature gravity \cite{Wald1993,IyerWald1994}. For a continuous interval of couplings, these two-theory relations describe the preferred saddles once minimization over the interval selects the endpoint theories.

Whether the preferred theories are the endpoints depends on the curvature interaction. We examine two families connected to Einstein--Maxwell gravity. The Gauss--Bonnet interaction is of Lovelock type and gives second-order metric equations. Its exact charged solution allows us to establish endpoint selection before minimizing over the horizon radius. The Weyl-squared interaction gives higher-derivative equations, so we use the perturbative charged branch \cite{CremoniniLiuSzepietowski2010,HuMaLuPang2024}. For an action with Weyl-squared as the sole correction, the free energy along each relevant stable thermal branch has negative coupling curvature at the Einstein point. This yields local endpoint selection near the Einstein transition. The two families also let us test how much of the canonical thermodynamics is fixed by anomaly data. Matching the anomaly coefficients gives the same first-order neutral Hawking--Page shift. At finite charge, matching the current normalization as well still leaves distinct small--large transition shifts.

Curvature-squared holography relates central charges to thermal transport \cite{KatsPetrov2009,BuchelMyersSinha2009}. Supersymmetric completions connect the same interactions to charged and rotating AdS black hole thermodynamics and anomaly matching \cite{CremoniniHanakiLiuSzepietowski2009,CassaniRuiperezTuretta2022}.

Our earlier work mapped charged Gauss--Bonnet transitions at fixed coupling \cite{Zeng:2025nzd}. We subsequently compared Gauss--Bonnet saddles at a fixed external source and showed how the cylinder energy prescription shifts the source required for five-dimensional endpoint coexistence \cite{Zeng2026Revisiting}. We adopt that potential and formulate the comparison covariantly by transforming the control together with the energy prescription. This framework gives the cross-CFT coexistence and slope-jump relations and lets us test endpoint selection in both Gauss--Bonnet and Weyl-squared gravity. We also give the conditions on the theory measure under which the classical minimum determines the leading free energy.

The rest of the paper follows the comparison from the boundary first law to the phase diagrams. Section~\ref{sec:framework} fixes the boundary data and derives the coexistence relations. Section~\ref{sec:families} gives the thermodynamic dictionaries and matches the anomaly and current data. Sections~\ref{sec:fixed-theory} and \ref{sec:phase-maps} treat the transitions within fixed CFTs and across their coupling intervals, respectively. Section~\ref{sec:discussion} discusses their physical and statistical interpretation. The appendices supply the radial, renormalization and higher-curvature calculations.

\section{Thermodynamic comparison on the boundary cylinder}
\label{sec:framework}
\label{sec:controls}

\begin{figure}[!htbp]
\centering
\includegraphics[width=.98\linewidth]{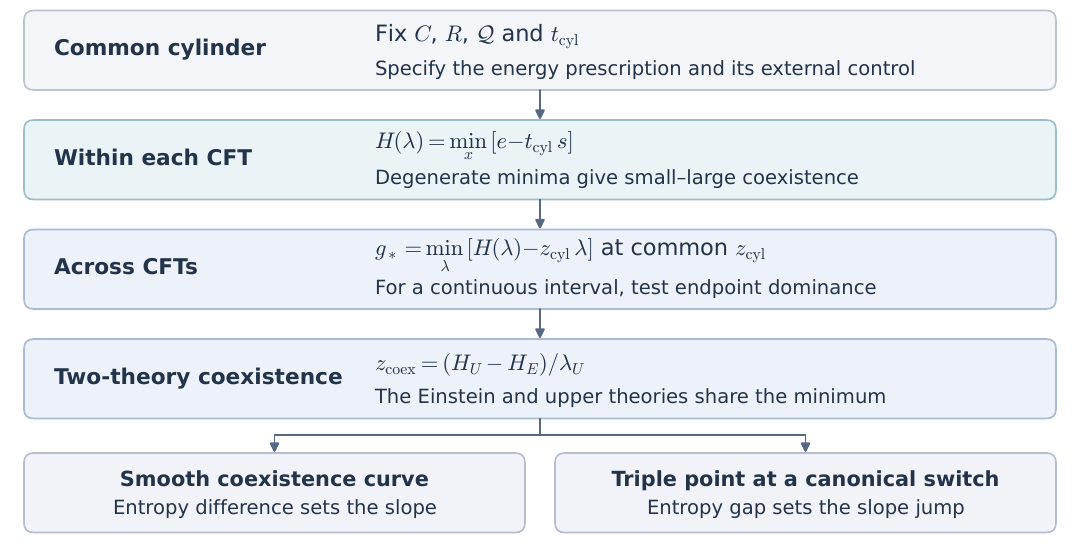}
\caption{Thermodynamic comparison within and across holographic CFTs. The common data are the overall normalization $C$, sphere radius $R$, conserved charge $\mathcal Q$ and reduced cylinder bath temperature $t_{\rm cyl}$. Here $x$ is the reduced horizon radius, $\lambda$ is the curvature coupling, and $e$ and $s$ are the reduced energy and entropy. Radial minimization gives the canonical minimum $H(\lambda)$. With the reference linear source in Eq.~\eqref{eq:grand}, minimizing $H(\lambda)-z_{\rm cyl}\lambda$ over couplings gives $g_*$. For the Einstein and upper-coupling pair, $\lambda_E=0$ and $\lambda_U>0$. Equation~\eqref{eq:cylinder-boundary} locates their coexistence. On a continuous interval this construction applies once an endpoint is shown to attain the global minimum. Equations~\eqref{eq:cylinder-clapeyron} and \eqref{eq:triple-crease} give the slope and its jump when one theory switches saddles.}
\label{fig:thermodynamic-framework}
\end{figure}

Figure~\ref{fig:thermodynamic-framework} summarizes how radial and coupling minimization connect canonical transitions to coexistence across theories.

\subsection{Boundary data and coupling work}

The fixed-coupling first law and the common boundary data suffice to derive the relations below. They apply to the exact Gauss--Bonnet branch and order by order to the Einstein-connected Weyl-squared branch with the additional higher-derivative boundary data held fixed. Section~\ref{sec:families} supplies the thermodynamic functions for these branches. Section~\ref{sec:phase-maps} gives the conditions for reducing an interval to its endpoints.

Consider a five-dimensional Einstein--Maxwell theory deformed by a dimensionless coupling $\lambda$. The boundary metric is $-dt_{\rm cyl,phys}^2+R^2d\Omega_3^2$, where $t_{\rm cyl,phys}$ is proper boundary time, $R$ is the sphere radius and $d\Omega_3^2$ is the unit round three-sphere metric. We quote extensive quantities per unit volume of the unit three-sphere. In both families we use
\begin{equation}
 C=\frac{L^3}{16\pi G_5},\quad x=\frac{r_+}{L},\quad
 \mathcal V=R^3,\quad \mathcal Q=CP,\quad P=2q,\quad b=\frac{q^2}{3}.
 \label{eq:coords}
\end{equation}
Here $L$ is the bare AdS length, $G_5$ is the five-dimensional Newton constant and $r_+$ is the areal horizon radius. The parameter $C$ sets the overall theory normalization and $\mathcal V$ is the reduced boundary volume. We use the same convention for the Maxwell flux $\mathcal Q$ throughout each coupling interval. The charges $P$ and $q$ are dimensionless, and $b$ is the metric charge parameter. At the Einstein point the electric term in the metric at the horizon is $b/x^4$. The comparison along each interval holds $C,R,\mathcal Q$ fixed.

The energy $E$ is measured in proper cylinder time. If the effective AdS radius is $L_{\rm eff}=\beta L$, then proper boundary time is $(R/L_{\rm eff})t_{\rm bulk}$, with $t_{\rm bulk}$ the asymptotic bulk time. The auxiliary $R/L$ clock uses bath and state temperatures $\Theta$ and $\widetilde T$, with reduced values $t=R\Theta$ and $\tau=R\widetilde T$. Their cylinder values are $\Theta_{\rm cyl}=\beta\Theta$ and $T_{\rm cyl}=\beta\widetilde T$. Let $S$ be the entropy and $\xi_{\rm cyl}$ the external source for the dimensionless coupling. The reduced cylinder variables are
\begin{equation}
 e=\frac{RE}{C},\quad s=\frac SC,\quad
 \tau_{\rm cyl}=RT_{\rm cyl},\quad t_{\rm cyl}=R\Theta_{\rm cyl},\quad
 z_{\rm cyl}=\frac{R\xi_{\rm cyl}}C .
 \label{eq:reduced}
\end{equation}

Subscripts on thermodynamic functions denote partial derivatives with respect to the indicated variable. The radial first law at fixed coupling gives $e_x=\tau_{\rm cyl}s_x$. When $\lambda$ varies at fixed entropy and charge, the radius must adjust to keep $s$ fixed. Combining that constraint with the radial first law gives
\begin{equation}
 \Xi_{\rm cyl}=\left(\frac{\partial E}{\partial\lambda}\right)_{S,\mathcal Q,C,\mathcal V}
 =\frac CR(e_\lambda-\tau_{\rm cyl}s_\lambda).
 \label{eq:xisource}
\end{equation}
Both derivatives on the right hold $x$ and $P$ fixed. For Gauss--Bonnet gravity the corresponding bulk variation enters the extended Lovelock first law \cite{KRT2010}. The cylinder electrostatic potential is $\Phi_{\rm cyl}$. Its reduced value $\phi=R\Phi_{\rm cyl}$ is conjugate to $P$ and satisfies $e_P=\tau_{\rm cyl}s_P+\phi$. In each family, $e$ and $s$ depend on $x,P,\lambda$. Differentiating $E=(C/R)e$ and using $S=Cs$ and $\mathcal Q=CP$ then yields
\begin{gather}
 dE=T_{\rm cyl}dS+\Phi_{\rm cyl}d\mathcal Q+
       \mu_CdC-p\,d\mathcal V+\Xi_{\rm cyl}d\lambda,\qquad
 p=\frac{E}{3\mathcal V},\label{eq:firstlaw}\\
 \mu_C=\frac{e-\tau_{\rm cyl}s-\phi P}{R},\qquad
 E=T_{\rm cyl}S+\Phi_{\rm cyl}\mathcal Q+\mu_C C .\label{eq:boundary-euler}
\end{gather}
The pressure is $p$, and $\mu_C$ is the response to varying $C$ at fixed entropy, charge, volume and coupling. The same work term can be expressed in a dimensionful coordinate $L_*^3\lambda$ with conjugate $\Xi_{\rm cyl}/L_*^3$. The reference length $L_*$ is common to all theories, so this coordinate is invariant under a rescaling of the bulk representative at fixed boundary data.

Charged rotating AdS black holes provide a complementary example of the role of boundary data. Varying their rotation parameters induces a boundary Weyl transformation. The resulting anomaly-induced variation of the Casimir energy enters the first law for the renormalized mass \cite{AwadElSayed2023}. In Eq.~\eqref{eq:xisource} the cylinder metric is fixed as the curvature coupling varies. The energy derivative includes any change in the cylinder vacuum energy along that theory path.

At fixed charge the canonical free energy is $\mathcal F=E-\Theta_{\rm cyl}S$. Including the common source term gives the potential $\mathcal G_{\rm cyl}$ and its reduced value $g_{\rm cyl}=R\mathcal G_{\rm cyl}/C$,
\begin{equation}
 \mathcal G_{\rm cyl}=E-\Theta_{\rm cyl}S-\xi_{\rm cyl}\lambda,
 \qquad g_{\rm cyl}=e-t_{\rm cyl}s-z_{\rm cyl}\lambda.
 \label{eq:grand}
\end{equation}
At fixed $\lambda$ the radial derivative is $(\tau_{\rm cyl}-t_{\rm cyl})s_x$. A thermal saddle has $\tau_{\rm cyl}=t_{\rm cyl}$, and it is radially stable when the potential has positive radial curvature at fixed bath data. Minimizing $e-t_{\rm cyl}s$ over the admissible radii defines $H(\lambda)$. The preferred potential $g_*$ is the minimum of $H(\lambda)-z_{\rm cyl}\lambda$ over candidate couplings. A smooth interior minimum in the coupling satisfies $\Xi_{\rm cyl}=\xi_{\rm cyl}$. At a radial stationary point on the lower or upper bound, the one-sided conditions are $\Xi_{\rm cyl}-\xi_{\rm cyl}\geq0$ and $\Xi_{\rm cyl}-\xi_{\rm cyl}\leq0$, respectively.

To assign probabilities across the family one must also choose a \emph{theory measure}, which weights discrete theories or intervals of a continuous family. The saddle action determines the leading exponential weight within each theory. Appendix~\ref{app:measure} gives the conditions on the additional theory measure under which the classical minimum determines the leading free energy.

We specify the energy prescription together with its external control. Our reference is holographic renormalization (HR) with zero coefficient for the type-D trace anomaly, which is proportional to the total derivative $\nabla^2\mathcal R_4$ of the boundary scalar curvature. The resulting vacuum energy is $E_{\rm vac}=3a_{\rm n}/(4R)$, or $e_{\rm vac}=RE_{\rm vac}/C$ in reduced units. Here $a$ is the Euler anomaly coefficient and $a_{\rm n}=a/(2\pi^2)$ is its value per unit sphere volume \cite{Assel2015,Benjamin2024}.

For a Gauss--Bonnet example of the source shift at fixed cylinder clock, see Appendix C of our previous study \cite{Zeng2026Revisiting}. In theory $j$, let $H_j=H(\lambda_j)$ be the minimum canonical free energy. Including the source term gives the potential $g_j=H_j-z_{\rm cyl}\lambda_j$ to be compared across theories. At fixed cylinder geometry, normalization and charge, changing the energy prescription adds a temperature-independent offset $\delta_j$ to $H_j$. For an Einstein and upper pair with $\lambda_E=0$ and $\lambda_U>0$, the transformation is
\begin{equation*}
 H_j^{\rm new}=H_j+\delta_j,\qquad
 z_{\rm cyl}^{\rm new}=z_{\rm cyl}+\frac{\delta_U-\delta_E}{\lambda_U},
 \qquad g_j^{\rm new}=g_j+\delta_E .
\end{equation*}
At the original source in one prescription and the transformed source in the other, both endpoint potentials acquire the same offset. This preserves their ordering and coexistence. At fixed charge, the source shift is independent of temperature, so it preserves the coexistence slope and its signed jumps. For the full coupling interval one compensates the shift as a function of coupling. Appendix~\ref{app:scheme-comparison} gives the general transformation and the condition for a linear control to suffice.

\subsection{Coexistence, entropy gaps and the role of the endpoints}
\label{sec:identities}

For the Einstein and upper pair, let $s_j$ be the reduced entropy of the saddle that attains $H_j$ on a smooth segment. At coexistence $H_U-z_{\rm cyl}\lambda_U=H_E$. The preferred potential $g_*$ and the coexistence source $z_{\rm coex}$ are
\begin{equation}
 g_*=\min\{H_E,H_U-z_{\rm cyl}\lambda_U\},\qquad
 z_{\rm coex}=\frac{H_U-H_E}{\lambda_U}.
 \label{eq:cylinder-boundary}
\end{equation}
The upper endpoint is preferred above the coexistence curve. Crossing it toward larger source changes the source derivative of $g_*$ by $-\lambda_U$. Along a smooth thermal branch, stationarity gives $\partial_{t_{\rm cyl}}H_j=-s_j$. Differentiating the equality of potentials therefore gives the fixed-charge Clapeyron relation \cite{WeiLiu2015Clapeyron},
\begin{equation}
 \frac{dz_{\rm coex}}{dt_{\rm cyl}}=-\frac{s_U-s_E}{\lambda_U}.
 \label{eq:cylinder-clapeyron}
\end{equation}
At a small--large transition in one endpoint theory, both of its black holes attain the same canonical minimum. They coexist with the preferred saddle of the other theory at the source in Eq.~\eqref{eq:cylinder-boundary}. Let $\Delta s_j>0$ denote the large-minus-small entropy gap at the switch. With primes denoting derivatives with respect to $t_{\rm cyl}$, the slope jump from the low- to the high-temperature side is
\begin{equation}
 [z_{\rm coex}']_{U}=-\frac{\Delta s_U}{\lambda_U},\qquad
 [z_{\rm coex}']_{E}=+\frac{\Delta s_E}{\lambda_U}.
 \label{eq:triple-crease}
\end{equation}
If both endpoints switch simultaneously, four saddles coexist and the two slope jumps add. At a canonical critical point the two radial branches merge, their entropy gap vanishes and the corresponding kink closes.

The two-endpoint formula applies on a continuous interval if the preferred value is always attained at one of the two endpoints. A sufficient condition is strict concavity in the coupling at each fixed radius on a common admissible domain. After radial minimization, the same argument applies if every thermal branch that attains the minimum at an interior coupling extends across the interval and is strictly concave there. At an interior coupling, the minimizing branch lies above the chord joining its endpoint values. Each of these values is at least as large as the canonical minimum at that endpoint. Subtracting the linear source term preserves the inequality. Applying these criteria to Gauss--Bonnet and Weyl-squared gravity requires their energy and entropy in the same cylinder convention.

\section{Two bulk families and their anomaly data}
\label{sec:families}

We obtain the canonical free energy and coupling response from the two bulk actions. Both Lorentzian actions have prefactor $(16\pi G_5)^{-1}$ and Lagrangian $\mathcal L$
\begin{equation}
 \mathcal L=\mathcal R+\frac{12}{L^2}-\frac14 F^2+
 \begin{cases}
 \frac12\lambda_{\rm GB}L^2\mathcal X_4,&\text{Gauss--Bonnet},\\
 \lambda_{\rm W}L^2 W_5^2,&\text{Weyl-squared}.
 \end{cases}
 \label{eq:actions}
\end{equation}
Here $F_{\mu\nu}$ is the Maxwell field strength, $F^2=F_{\mu\nu}F^{\mu\nu}$, and curvature squares denote full tensor contractions. The labels $\mathrm{GB}$ and $\mathrm{W}$ refer to the Gauss--Bonnet and Weyl-squared families. The quantity $\mathcal X_4=\mathcal R_{\mu\nu\rho\sigma}^2-4\mathcal R_{\mu\nu}^2+\mathcal R^2$ is the Gauss--Bonnet density. The quantity $W_5^2$ is the square of the five-dimensional Weyl tensor. The general coupling $\lambda$ denotes the corresponding parameter of the chosen family. We impose standard AdS boundary conditions on the branches connected to Einstein--Maxwell gravity. The Weyl-squared calculation is perturbative, with the independent boundary data of the additional higher-derivative modes set to zero. Within the smooth spherical saddle family, a nonzero Maxwell flux in the absence of charged matter requires a charged black hole.

Using the reduced variables of Sec.~\ref{sec:framework}, the canonical free energy, pressure and normalization conjugate on a thermal saddle are
\begin{equation*}
 \mathcal F_{\rm on}=\frac{C}{R}(e-\tau_{\rm cyl}s),\qquad
 p=\frac{Ce}{3R^4},\qquad R\mu_C=e-\tau_{\rm cyl}s-\phi P.
\end{equation*}

\subsection{Gauss--Bonnet dictionary}

In the Gauss--Bonnet formulas and figure axes, $\lambda=\lambda_{\rm GB}$. With $r$ the areal radial coordinate and $f_{\rm GB}$ the spherical metric function, define $\psi=(1-f_{\rm GB})/r^2$ and $\alpha=\lambda_{\rm GB}L^2$. The spherical metric obeys the following equation, with $M$ the mass relative to AdS \cite{Cai2013},
\begin{equation}
 \psi+\alpha\psi^2=\frac{m_{\rm bulk}}{r^4}-\frac1{L^2}-\frac{bL^4}{r^6},
 \qquad m_{\rm bulk}=\frac{16\pi G_5}{3}M .
\end{equation}
The Einstein root has $f_{\rm GB}=1+\allowbreak r^2[1-\sqrt{1+4\alpha(m_{\rm bulk}/r^4-L^{-2}-bL^4/r^6)}]/(2\alpha)$. At the horizon $\psi=r_+^{-2}$, so $m_{\rm bulk}/L^2=m+\lambda_{\rm GB}$, where $m(x,b)=x^4+x^2+b/x^2$ is the dimensionless Einstein mass function. Setting the mass and flux to zero gives the effective AdS radius. For the entropy, the Wald functional reduces to the Lovelock horizon functional \cite{JacobsonMyers1993,IyerWald1994}. With horizon scalar curvature $6/r_+^2$, these results give
\begin{gather}
 \eta=\sqrt{1-4\lambda_{\rm GB}},\qquad
 \beta=\frac{L_{\rm eff}}L=\sqrt{\frac{1+\eta}{2}},\label{eq:ads-scale}\\
 s_{\rm GB}=4\pi(x^3+6\lambda_{\rm GB}x).\label{eq:entropy}
\end{gather}
Background subtraction (BS) assigns zero energy to pure AdS at each coupling \cite{HawkingHorowitz1996}. In the $R/L$ clock the reduced energy is $e_{R/L}^{\rm BS}=RE_{R/L}^{\rm BS}/C=3(m+\lambda_{\rm GB})$. The surface gravity gives the reduced state temperature in the same clock, in agreement with the radial first law,
\begin{equation}
 \tau=\frac{6x^6+3x^4-q^2}{6\pi x^3(x^2+2\lambda_{\rm GB})},
 \qquad \tau_{\rm cyl}=\beta\tau .
 \label{eq:temperature}
\end{equation}
To obtain the cylinder energy in the zero-type-D prescription, we rescale the mass contribution by $\beta$ and add the vacuum energy,
\begin{equation}
 e_{\rm GB}=3\beta(m+\lambda_{\rm GB})+e_{\rm vac}^{\rm GB}(\lambda_{\rm GB}),\qquad
 e_{\rm vac}^{\rm GB}=\frac34\beta^3(3\eta-2).
 \label{eq:proper-time}
\end{equation}
The electrostatic potential in the same cylinder clock is
\begin{equation*}
 \phi_{\rm GB}=\frac{\beta P}{2x^2}.
\end{equation*}
Appendix~\ref{app:stress} obtains the energy from the boundary stress tensor and checks the radial and electric parts of the first law against the horizon quantities. Substituting the cylinder energy and Wald entropy into the fixed-entropy derivative in Eq.~\eqref{eq:xisource} gives
\begin{equation}
 \frac{R\Xi_{\rm cyl}^{\rm GB}}C
 =\beta\left(3-24\pi x\tau-\frac{45}{8}\right)
 -\frac{3(m+\lambda_{\rm GB})}{2\eta\beta}.
 \label{eq:gb-work}
\end{equation}
The vacuum term uses $\partial_{\lambda_{\rm GB}}e_{\rm vac}^{\rm GB}=-45\beta/8$. The clock term uses $\partial_{\lambda_{\rm GB}}\beta=-1/(2\eta\beta)$.

\subsection{Weyl-squared dictionary}

The Weyl tensor vanishes in pure AdS, so $L_{\rm eff}=L$ along the Weyl-squared path and the two time conventions coincide. For the charged solution use $\rho=r/L$ and $\widehat t=t_{\rm bulk}/L$. In these dimensionless areal coordinates, the metric is
\begin{equation}
 \frac{ds^2}{L^2}=-B_t(\rho)d\widehat t^2+
 \frac{d\rho^2}{f_{\rm W}(\rho)}+\rho^2d\Omega_3^2,\qquad B_t=N^2f_{\rm W} .
 \label{eq:weyl-metric}
\end{equation}
Here $f_{\rm W}$ is the radial metric function, $N$ the lapse and $B_t$ the time component factor. At zero coupling the metric reduces to Reissner--Nordstr\"om--AdS, with $B_t=f_{\rm W}=1+\rho^2-\mu/\rho^2+b/\rho^4$. Imposing a horizon at $x$ fixes the dimensionless mass parameter $\mu=m+\lambda_{\rm W}m_1+O(\lambda_{\rm W}^2)$, where
\begin{equation*}
 m_1=-\frac{40b}{x^2}+\frac{2m^2-20b}{x^4}+\frac{4mb}{x^6}+\frac{5b^2}{4x^8}.
\end{equation*}
For $Y\in\{e,s,\tau_{\rm cyl},\phi\}$, write $Y=Y_0+\lambda_{\rm W}Y_1+O(\lambda_{\rm W}^2)$. At fixed areal radius and charge, the zeroth-order coefficients are
\begin{equation}
 e_0=3m+\frac34,\quad s_0=4\pi x^3,\quad
 \tau_0=\frac{2x+x^{-1}-bx^{-5}}{2\pi},\quad
 \phi_0=\frac{P}{2x^2}.
 \label{eq:weyl-thermo-zero}
\end{equation}
The first-order coefficients at the same radius and charge are
\begin{equation}
 \begin{gathered}
  e_1=3m_1+24m,\qquad
  s_1=4\pi x^3\left(\frac{12m}{x^4}-\frac{30b}{x^6}\right),\\
  \tau_1=-\frac{11b^2-10bx^4+4x^8(1+x^2)^2}{2\pi x^{11}},\qquad
  \phi_1=\frac{5P^3}{48x^8}.
 \end{gathered}
 \label{eq:weyl-thermo-linear}
\end{equation}
The black hole energy follows from the time component of the renormalized boundary stress tensor. Let $\widehat\alpha_1,\widehat\alpha_2,\widehat\alpha_3$ be the coefficients of the scalar-curvature square, Ricci square and Gauss--Bonnet density after dividing out $L^2$. The stress-tensor calculation of Cremonini, Liu and Szepietowski gives the following mass relative to AdS, written in five dimensions in our reduced units \cite{CremoniniLiuSzepietowski2010} (their Eqs.~(53) and (54)),
\begin{equation*}
 e_{\rm W}^{\rm BS}
 =3\mu\left(1-40\widehat\alpha_1-8\widehat\alpha_2-4\widehat\alpha_3\right)
  +O(\lambda_{\rm W}^2).
\end{equation*}
For the Weyl-squared interaction the three coefficients are $(-5/6,8/3,1)\lambda_{\rm W}$. The correction in the parentheses is $(100/3-64/3-4)\lambda_{\rm W}=8\lambda_{\rm W}$. Adding the cylinder vacuum energy and expanding the horizon mass parameter then gives
\begin{equation*}
 e=\tfrac34+3(1+8\lambda_{\rm W})\mu+O(\lambda_{\rm W}^2)
   =3m+\tfrac34+\lambda_{\rm W}(3m_1+24m)+O(\lambda_{\rm W}^2).
\end{equation*}
Appendix~\ref{app:weyl} derives the horizon quantities, verifies the first law and independently recovers the energy correction from the fixed-charge action.

The fixed-entropy coupling response at the Einstein point is
\begin{equation}
 \left.\frac{R\Xi_{\rm cyl}^{\rm W}}C\right|_0=e_1-\tau_0s_1=w(x,b),
 \quad
 w=-\frac{18m^2}{x^4}+\frac{60mb}{x^6}-\frac{225b^2}{4x^8}.
 \label{eq:weyl-work}
\end{equation}
The response can also be obtained from the explicit higher-derivative insertion on the unperturbed saddle \cite{ReallSantos2019}. At fixed bath temperature and charge, the field variation of the two-derivative action vanishes by stationarity \cite{CremoniniJonesLiuMcPeak2020}. The remaining Weyl-squared integral is
\begin{equation}
 w(x,b)=-\int_x^\infty d\rho\,\rho^3 L^4W_{5,0}^2,\qquad
 L^4W_{5,0}^2=\frac12\left(-\frac{12m}{\rho^4}+\frac{30b}{\rho^6}\right)^2.
 \label{eq:weyl-insertion}
\end{equation}
On the round cylinder the explicit surface contribution vanishes \cite{HuMaLuPang2024,MaHuPangLu2024}, as shown in Appendix~\ref{app:weyl}. As a check, the charged AdS grand potential of Hu et al.\ \cite{HuMaLuPang2024} (their Eqs.~(90)--(92)) gives the following reduced first-order coefficient for a five-dimensional spherical horizon and the Weyl-squared term alone,
\begin{equation*}
 -18(1+x^2)^2+\frac{24b(1+x^2)}{x^4}-\frac{57b^2}{4x^8}.
\end{equation*}
In taking the coupling derivative of the Legendre transform at fixed charge, the derivative of the potential cancels by stationarity. The grand-potential coefficient is thus evaluated at the fixed-charge Einstein saddle \cite{CremoniniJonesLiuMcPeak2020}. Substituting $m/x^2=1+x^2+b/x^4$ in Eq.~\eqref{eq:weyl-work} reproduces the displayed coefficient. The sign of the integral in Eq.~\eqref{eq:weyl-insertion} shows that the interaction lowers the canonical free energy on each smooth charged Einstein saddle at fixed temperature and charge.

For the continuation with the Weyl-squared term as the sole correction, the response through linear order at fixed areal radius is
\begin{equation*}
 \frac{R\Xi_{\rm cyl}^{\rm W}}{C}=w(x,b)+\lambda_{\rm W}K_{\rm ins}(x,b)+O(\lambda_{\rm W}^2).
\end{equation*}
The coefficient $K_{\rm ins}$ comes from varying the insertion on the first-order geometry at fixed radius and charge, as in Eq.~\eqref{eq:weyl-insertion-linear}. This term enters the thermal-branch curvature used for local endpoint selection in Sec.~\ref{sec:weyl-selection}.

\subsection{Comparing the paths in anomaly space}
\label{sec:anomaly}

The dictionaries also define trajectories in anomaly space. The logarithmic term in the bulk action determines the curvature-squared contributions to the boundary trace anomaly \cite{NojiriOdintsov2000,FukumaMatsuuraSakai2001}. We use $a$ and $c$ for the Euler and Weyl-squared coefficients and divide each by $2\pi^2$ to obtain $a_{\rm n}$ and $c_{\rm n}$.

For Gauss--Bonnet gravity the anomaly coefficients of Myers and Singh \cite{MyersSingh2012} (their Eq.~(5.8)) read
\begin{equation*}
 (a,c)=\frac{\pi L_{\rm eff}^3}{8G_5}
 \left(1-6\lambda_{\rm GB}f_\infty,\;1-2\lambda_{\rm GB}f_\infty\right).
\end{equation*}
Here $f_\infty=\beta^{-2}$ and $2\lambda_{\rm GB}f_\infty=1-\eta$. For the Weyl-squared path the anomaly formula of Ma et al.\ \cite{MaHuPangLu2024} (their Eq.~(10)) reduces to
\begin{equation*}
 (a,c)=\frac{\pi L^3}{8G_5}\left(1,\;1+\frac{8c_3}{L^2}\right)+O(\lambda_{\rm W}^2),
 \qquad c_3=L^2\lambda_{\rm W}.
\end{equation*}
The radius is unchanged because the Weyl tensor vanishes on pure AdS. Substituting these identifications and dividing by the unit three-sphere volume gives
\begin{align}
 (a_{\rm n}^{\rm GB},c_{\rm n}^{\rm GB})&=C\beta^3(3\eta-2,\eta),\label{eq:gb-anomalies}\\
 (a_{\rm n}^{\rm W},c_{\rm n}^{\rm W})&=C(1,1+8\lambda_{\rm W})
       +O(\lambda_{\rm W}^2).\label{eq:w-anomalies}
\end{align}
The first-order Weyl-squared logarithmic term on a general boundary is $C\lambda_{\rm W}W_4^2$, with $W_4$ its Weyl tensor. Consequently, only $c_{\rm n}$ changes at this order. In Gauss--Bonnet gravity the Jacobian from $(C,\lambda_{\rm GB})$ to $(a_{\rm n},c_{\rm n})$ is $C(1+\eta)^3/(2\eta)$. Both families have Jacobian $8C$ at the Einstein point. At fixed $C$ their tangents there are $C(-15/2,-7/2)$ and $C(0,8)$, respectively. The varying anomaly coefficients distinguish the CFTs along each fixed-$C$ path.

The ratios $a/c$ are $3-2/\eta$ for Gauss--Bonnet and $1-8\lambda_{\rm W}+O(\lambda_{\rm W}^2)$ for Weyl-squared gravity. Expanding the Gauss--Bonnet pair near the Einstein point gives
\begin{equation*}
 \frac{(a_{\rm n}^{\rm GB},c_{\rm n}^{\rm GB})}{C_{\rm GB}}
  =\left(1-\tfrac{15}{2}\lambda_{\rm GB},\;1-\tfrac72\lambda_{\rm GB}\right)
    +O(\lambda_{\rm GB}^2).
\end{equation*}
Equating the ratio $a/c$ to the Weyl-squared ratio first fixes the relative couplings. Equating the Euler coefficients then fixes the Gauss--Bonnet normalization relative to the Weyl-squared normalization $C$,
\begin{equation}
 \lambda_{\rm GB}=2\lambda_{\rm W}+O(\lambda_{\rm W}^2),
 \qquad C_{\rm GB}=C(1+15\lambda_{\rm W})+O(\lambda_{\rm W}^2).
 \label{eq:anomaly-match}
\end{equation}

For the charged comparison we also match the normalization of the conserved current. In a flat boundary patch the current two-point function calculated by Freedman et al.\ \cite{FreedmanMathurMatusisRastelli1999} (their Eq.~(54)) has the four-dimensional form
\begin{equation*}
 \langle J_i(X)J_j(0)\rangle
 =\frac{C_J}{|X|^6}\left(\delta_{ij}-2\frac{X_iX_j}{|X|^2}\right).
\end{equation*}
Here $X$ is a Euclidean boundary separation, and $C_J$ is proportional to the coefficient of the bulk Maxwell action in unit-radius AdS. Restoring the radius in five bulk dimensions contributes a factor of $L_{\rm eff}$. Thus $C_J\propto ZL_{\rm eff}/G_5$, where $Z$ multiplies the Maxwell term. Equivalently, in Poincar\'e coordinates the radial Maxwell kinetic coefficient is proportional to $ZL_{\rm eff}/(G_5z_{\rm FG})$, where $z_{\rm FG}$ is the Fefferman--Graham coordinate. At common bare $L$ and current-source unit, equality of the current coefficients and of the physical flux gives
\begin{equation*}
 Z_{\rm GB}\,\beta\,\frac{C_{\rm GB}}C=1,\qquad
 C_{\rm GB}\sqrt{12Z_{\rm GB}b_{\rm GB}}=C\sqrt{12b}.
\end{equation*}
Using Eq.~\eqref{eq:anomaly-match} and $\beta=1-\lambda_{\rm W}+O(\lambda_{\rm W}^2)$ gives
\begin{equation}
 Z_{\rm GB}=1-14\lambda_{\rm W}+O(\lambda_{\rm W}^2),\qquad
 b_{\rm GB}=b(1-16\lambda_{\rm W})+O(\lambda_{\rm W}^2).
 \label{eq:charge-match}
\end{equation}
These rescalings are specific to the comparison at matched data. The cross-CFT phase diagrams vary the curvature coupling within each family at its original Maxwell normalization.

Measure both free energies in units of $C/R$, with $C$ the common Einstein normalization. At fixed cylinder temperature the linear change of the matched Gauss--Bonnet free energy is $15h_0-16b\,\partial_b h_0+2B_{\rm GB}$. Here $h_0$ is the reduced Einstein free energy at fixed radius and $B_{\rm GB}$ is the coupling derivative at fixed radius and metric charge, both evaluated at the common Einstein point.
\begin{gather}
 h_0=3m+\tfrac34-4\pi t_{\rm cyl}x^3,\qquad
 B_{\rm GB}=3-\tfrac32m-\tfrac{45}{8}-24\pi t_{\rm cyl}x .
\end{gather}
The terms arise from the rescalings of $C$, the metric charge and the Gauss--Bonnet coupling. Radial stationarity removes the contribution from the radius shift. Evaluating the difference on the Einstein saddle with $t_{\rm cyl}=\tau_0(x,b)$ yields
\begin{equation}
 \frac{R}{C}\left(\mathcal F_{\rm W}-\mathcal F_{\rm GB}^{\rm matched}\right)
 =-\lambda_{\rm W}\frac{57b^2}{4x^8}+O(\lambda_{\rm W}^2).
 \label{eq:matched-free}
\end{equation}
Here $x$ is the common zeroth-order saddle radius. At fixed zeroth-order radius, the leading difference is quartic in the flux and vanishes at zero flux. Curvature field redefinitions can induce additional Maxwell interactions in Einstein--Maxwell theory \cite{LiuSzepietowski2009}, so equal anomaly and current data still allow different finite-charge responses. Section~\ref{sec:matched-coexistence} uses this free-energy difference to compare the canonical transition temperatures.

\section{Canonical transitions within fixed CFTs}
\label{sec:fixed-theory}

The energies and entropies of Sec.~\ref{sec:families} determine the canonical minimum at each fixed coupling. In comparing the small and large saddles in Eq.~\eqref{eq:grand}, the source term and vacuum energy cancel at that coupling. This gives the canonical transition whose entropy gap enters the cross-CFT comparison.
\subsection{Gauss--Bonnet gravity}
For Gauss--Bonnet gravity it is convenient to compute the canonical transition in the $R/L$ clock. To convert to cylinder units, multiply both the temperature and the free-energy difference by $\beta(\lambda)>0$. At nonzero charge the spherical saddles are black holes. If the state temperature has two turning points, their temperatures bound an interval with three saddles. These turning points are the spinodals, where a stable and an unstable radial branch merge. The transition occurs where the two stable minima have equal free energy \cite{Chamblin1999,Cong2022,Lyu2024}.

\begin{figure}[tbp]
\centering
\includegraphics[width=.98\linewidth]{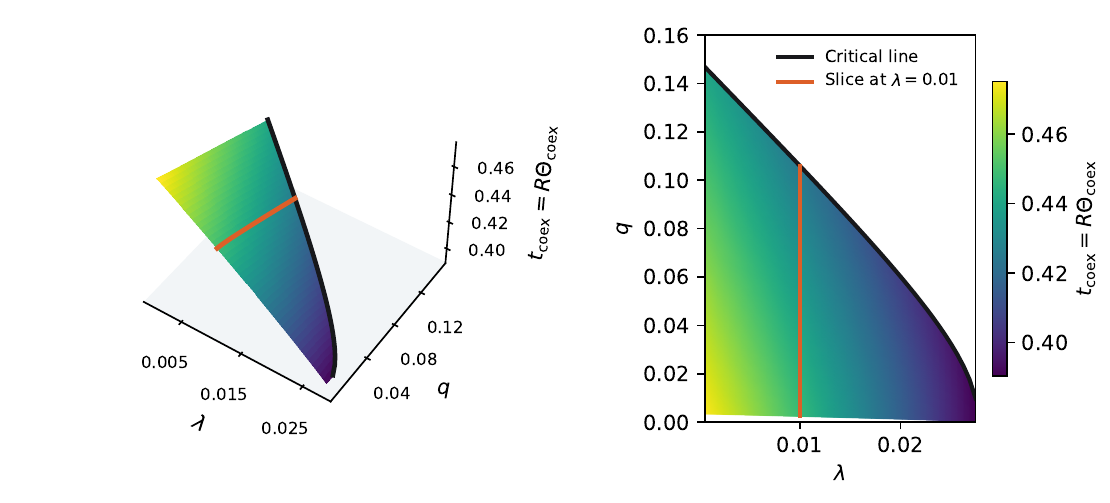}
\caption{Gauss--Bonnet canonical coexistence at fixed $C,R$. The left panel shows the surface and the right panel projects it onto the plane of the Gauss--Bonnet coupling $\lambda$ and reduced charge $q$. Height and color give the reduced coexistence temperature $t_{\rm coex}$ in the $R/L$ clock. Multiplication by $\beta(\lambda)=L_{\rm eff}/L$ gives the cylinder value. Each point solves Eq.~\eqref{eq:canonical-coex} at fixed $\lambda,q$. The black curve is the critical edge given by Eq.~\eqref{eq:canonical-critical-edge}, and the orange curve marks $\lambda=0.01$. Equation~\eqref{eq:triple-param} gives the Einstein slice. The displayed range is $0.0005\leq\lambda\leq0.0275$ and $0.025q_c(\lambda)\leq q\leq q_c(\lambda)$.}
\label{fig:charged-surface}
\end{figure}

For Fig.~\ref{fig:charged-surface} the reduced potential in the $R/L$ clock is
\begin{equation}
 h_{\rm GB}^{R/L}(x;t,\lambda,q)=3(m+\lambda)-t\,s_{\rm GB}(x,\lambda),
 \label{eq:fig-canonical-potential}
\end{equation}
This uses the background-subtracted energy, with $m=m(x,b)$ and $s_{\rm GB}$ from Eq.~\eqref{eq:entropy}. At bath temperature $t$, Eq.~\eqref{eq:temperature} gives the stable radii $x_s<x_l$ (small and large, respectively) through $\tau(x_s)=\tau(x_l)=t$. The coexistence temperature $t_{\rm coex}$ solves $h_{\rm GB}^{R/L}(x_l)-h_{\rm GB}^{R/L}(x_s)=0$. Since $\partial_x h_{\rm GB}^{R/L}=(\tau-t)s_x$ and $s_x>0$, this becomes an equal-area condition between the reduced entropies $s_s=s_{\rm GB}(x_s,\lambda)$ and $s_l=s_{\rm GB}(x_l,\lambda)$. At fixed $\lambda,q$,
\begin{equation}
 \tau(s_s)=\tau(s_l)=t_{\rm coex},\qquad \int_{s_s}^{s_l}[\tau(s)-t_{\rm coex}]\,ds=0 .
 \label{eq:canonical-coex}
\end{equation}
The integral gives the large-minus-small free-energy difference in units of $C/R$. It crosses zero once between the spinodals, as shown in Appendix~\ref{app:repro}. For $0<\lambda<1/36$, set $y=x^2$ and let $y_-<y_+$ be the positive roots of $2y^2+(12\lambda-1)y+2\lambda$. The turning points satisfy $q^2=W_{\rm sp}(y;\lambda)$, where the spinodal charge function obeys $W_{\rm sp}=6y^2(y-y_-)(y_+-y)/(5y+6\lambda)$ for $y_-<y<y_+$. The spinodals merge at the unique maximum $y_c(\lambda)$ of $W_{\rm sp}$. This defines the critical charge and the reduced critical temperature in the $R/L$ clock,
\begin{equation}
\partial_y W_{\rm sp}(y_c;\lambda)=0,\quad
q_c(\lambda)=\sqrt{W_{\rm sp}(y_c;\lambda)},\quad
t_c(\lambda)=\tau\big(\sqrt{y_c};\lambda,q_c(\lambda)\big).
\label{eq:canonical-critical-edge}
\end{equation}
Where $W_{\rm sp}$ is positive, increasing $\lambda$ decreases its numerator $-3y^2[2y^2+(12\lambda-1)y+2\lambda]$ and increases its denominator. Evaluating the two functions at the maximum for the larger coupling shows that $q_c$ decreases with $\lambda$. As $\lambda$ approaches $1/36$, the roots $y_-$ and $y_+$ merge at $y=1/6$ and $q_c$ tends to zero. At zero charge smooth thermal AdS also enters the canonical comparison. The limiting black hole lies at the neutral Hawking--Page temperature. Substitution of this radius and coupling into $E_{R/L}^{\rm BS}-\widetilde T S$ gives the thermal AdS free energy of zero in this prescription \cite{HawkingPage1983,Cai2002,Witten1998}. The entropy of the limiting black hole is finite and gives the entropy jump across the neutral transition.

For the displacement of the transition at fixed charge, the explicit coupling derivative of Eq.~\eqref{eq:fig-canonical-potential} is $3-24\pi t x$. Radial stationarity removes the radius derivatives from the on-shell free-energy difference. At coexistence its coupling derivative is $-24\pi t_{\rm coex}(x_l-x_s)$ and its temperature derivative is $-(s_l-s_s)$. Differentiating the coexistence condition gives
\begin{equation}
\frac{dt_{\rm coex}}{d\lambda}
=-24\pi t_{\rm coex}\frac{x_l-x_s}{s_l-s_s}<0 .
\label{eq:canonical-coex-slope}
\end{equation}
The reduced cylinder temperature of the Gauss--Bonnet small--large transition is $t_{\rm sl}^{\rm GB}=\beta(\lambda)t_{\rm coex}$. Both factors decrease with coupling. Thus, whenever both endpoint theories pass through the canonical transition, the upper Gauss--Bonnet endpoint switches first in the cylinder clock. The decrease in $q_c(\lambda)$ also implies that coexistence in both endpoint theories at a given charge extends to all intermediate slices.

For Fig.~\ref{fig:charged-surface}, we sample $q/q_c(\lambda)$ on each fixed-coupling slice. The two spinodal temperatures bracket the free-energy crossing, as described in Appendix~\ref{app:repro}.

At fixed coupling, radial stationarity leaves only the explicit charge derivative of the free-energy difference, $2q(x_l^{-2}-x_s^{-2})$. Dividing by the entropy gap gives
\begin{equation}
 \left.\frac{\partial t_{\rm coex}}{\partial q}\right|_{\lambda}
 =\frac{2q(x_l^{-2}-x_s^{-2})}{s_l-s_s}<0.
 \label{eq:canonical-charge-slope}
\end{equation}
Increasing either parameter lowers the bath temperature at which the large black hole becomes preferred. At the black edge the two radii merge, so the entropy gap and latent heat vanish. The decrease in $q_c$ narrows the range of charges with a first-order transition. This edge will determine whether the upper Gauss--Bonnet theory contributes a second kink to the cross-CFT curve in Sec.~\ref{sec:phase-maps}.

At $q=1/(16\pi)$, the coexistence temperature falls from $t_{\rm coex}\simeq0.45881480$ at $\lambda=0.005$ to $0.44465777$ at $\lambda=0.01$. Solving $q=q_c(\lambda)$ for the critical coupling at this charge gives $\lambda_c\simeq0.02665899$ and $t_c\simeq0.39241425$.

\subsection{Weyl-squared gravity}

The Weyl-squared calculation follows the canonical transition perturbatively from the same Einstein limit. Related fixed-theory phase transitions have been studied for the curvature--Maxwell interaction $W_{\mu\nu\rho\sigma}F^{\mu\nu}F^{\rho\sigma}$ \cite{Mahapatra2016}.

At a fixed cylinder bath temperature let $x_0$ be a nondegenerate root of $\tau_0(x_0,b)=t_{\rm cyl}$. With primes denoting radial derivatives at fixed $b$, Eqs.~\eqref{eq:weyl-thermo-zero} and \eqref{eq:weyl-thermo-linear} give the radius and entropy of the perturbed saddle,
\begin{equation}
 x=x_0+\lambda_{\rm W}x_1+O(\lambda_{\rm W}^2),\qquad
 x_1=-\frac{\tau_1(x_0,b)}{\tau_0'(x_0,b)},\qquad
 s_{\rm on}=s_0+\lambda_{\rm W}(s_1+s_0'x_1)+O(\lambda_{\rm W}^2).
 \label{eq:weyl-shifted-saddle}
\end{equation}
The radius shift enters the entropy explicitly. Its contribution to the first-order free energy vanishes because $e_0-t_{\rm cyl}s_0$ is stationary at $x_0$. The free-energy shift is then the insertion evaluated at that radius,
\begin{equation}
 h_{\rm on}=e_0(x_0,b)-t_{\rm cyl}s_0(x_0)+\lambda_{\rm W}w(x_0,b)
               +O(\lambda_{\rm W}^2).
 \label{eq:weyl-onshell}
\end{equation}
Differentiating Eq.~\eqref{eq:weyl-onshell} with respect to the bath temperature at fixed charge gives minus the Wald entropy in Eq.~\eqref{eq:weyl-shifted-saddle}, including the term from the displaced radius.

Let $t_{\rm sl}^{\rm W}$ be the reduced Weyl-squared small--large coexistence temperature on the cylinder. At a fixed subcritical charge with distinct stable saddles, the temperature derivative of the large-minus-small free energy at the Einstein crossing is $-\Delta s_E$, where $\Delta s_E$ is the reduced entropy gap. The coupling derivative is $w_l-w_s$, with insertions evaluated at the large and small Einstein radii. Differentiating the equality of free energies gives
\begin{equation}
 \left.\frac{dt_{\rm sl}^{\rm W}}{d\lambda_{\rm W}}\right|_0
  =\frac{w_l-w_s}{\Delta s_E}.
 \label{eq:weyl-switch}
\end{equation}
The sign follows from the Einstein coexistence parameterization in Eq.~\eqref{eq:triple-param}, with $k=x_s/x_l$ and denominator polynomial $D_k$. Inserting the two radii into Eq.~\eqref{eq:weyl-work} gives
\begin{gather}
 w_l-w_s=-\frac{3(1-k^2)}{4D_k^2}P_6(k),\\
 P_6(k)=85k^6+528k^5+1373k^4+1872k^3+1373k^2+528k+85 .
\end{gather}
For $0<k<1$, both the polynomial and the entropy gap are positive. Hence a sufficiently small positive Weyl-squared coupling lowers the canonical transition temperature at fixed subcritical charge.

At the critical edge the two radii merge and the nondegenerate radius shift in Eq.~\eqref{eq:weyl-shifted-saddle} ceases to apply. Instead the critical state satisfies $\partial_x\tau_{\rm cyl}=\partial_x^2\tau_{\rm cyl}=0$ at fixed flux. Expanding these conditions with $\tau_{\rm cyl}=\tau_0+\lambda_{\rm W}\tau_1+O(\lambda_{\rm W}^2)$ from Eqs.~\eqref{eq:weyl-thermo-zero} and \eqref{eq:weyl-thermo-linear} gives the shifts from the Einstein critical values $q_{{\rm c},E}=1/\sqrt{45}$ and $t_{{\rm c},E}=4\sqrt{3}/(5\pi)$,
\begin{gather}
 q_c^{\rm W}=q_{{\rm c},E}+\lambda_{\rm W}q_{c,1}+O(\lambda_{\rm W}^2),\qquad
 q_{c,1}=-\frac{757\sqrt{5}}{250},\notag\\
 t_c^{\rm W}=t_{{\rm c},E}+\lambda_{\rm W}t_{c,1}+O(\lambda_{\rm W}^2),\qquad
 t_{c,1}=-\frac{839\sqrt{3}}{125\pi}.
 \label{eq:weyl-critical-edge}
\end{gather}

To follow the moving critical edge, we expand at fixed fraction $\varrho=q/q_c^{\rm W}$. Let $q_{\rm ref}=\varrho q_{{\rm c},E}$ and take $t_{{\rm sl},E}(q_{\rm ref})$ from the Einstein coexistence curve in Eq.~\eqref{eq:triple-param}. The resulting temperature at $q=\varrho q_c^{\rm W}$ is
\begin{align}
 t_{\rm sl}^{\rm W}
 &=t_{{\rm sl},E}(q_{\rm ref})+\lambda_{\rm W}\left[
     \frac{w_l-w_s}{\Delta s_E}
     +\varrho q_{c,1}\left.\frac{dt_{{\rm sl},E}}{dq}\right|_{q_{\rm ref}}\right]
     +O(\lambda_{\rm W}^2).
 \label{eq:weyl-canonical-surface}
\end{align}
The insertions and entropy gap are evaluated at the Einstein coexistence radii for $q_{\rm ref}$. The second term in brackets accounts for the change in charge at fixed $\varrho$. The numerators and denominators of the ratios vanish together at $k=1$. Cancelling the common factor $1-k$ gives a finite limit at $\varrho=1$ and gives $t_{c,1}$ in Eq.~\eqref{eq:weyl-critical-edge}. Appendix~\ref{app:weyl-canonical} derives the critical shift and the regular expansion at the edge.

\begin{figure}[tbp]
\centering
\includegraphics[width=.98\linewidth]{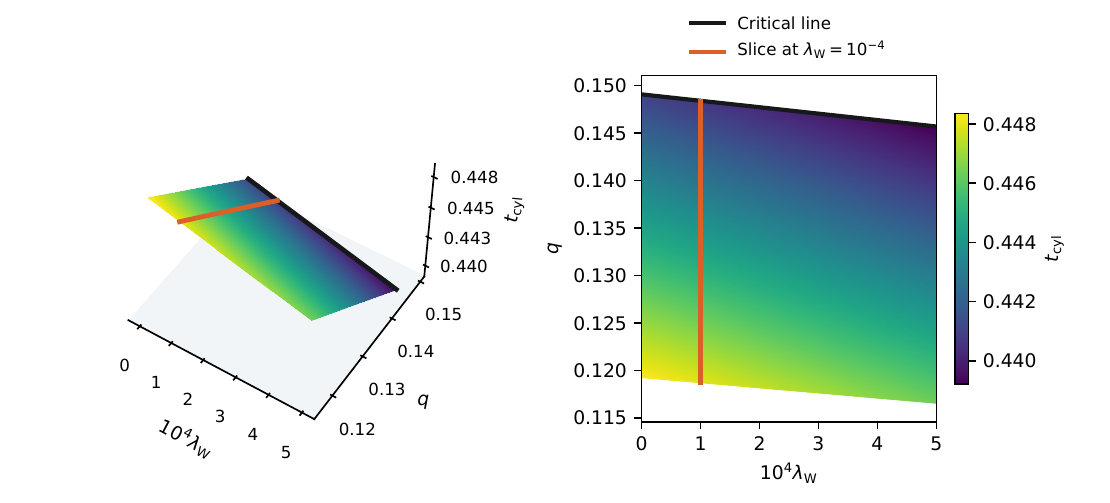}
\caption{Weyl-squared canonical coexistence at fixed $C,R$. The left panel shows the surface and the right panel projects it onto the plane of the Weyl-squared coupling $\lambda_{\rm W}$ and reduced charge $q$. Height and color give the reduced cylinder temperature at which the small and large black holes have equal free energies, from Eq.~\eqref{eq:weyl-canonical-surface}. The black curve is the critical edge in Eq.~\eqref{eq:weyl-critical-edge}, and the orange curve marks $\lambda_{\rm W}=10^{-4}$. The first-order surface is displayed for $0\leq\lambda_{\rm W}\leq5\times10^{-4}$ and $0.8q_c^{\rm W}\leq q\leq q_c^{\rm W}$.}
\label{fig:weyl-canonical-surface}
\end{figure}

To plot the surface in Fig.~\ref{fig:weyl-canonical-surface}, we sample $\varrho$ at each $\lambda_{\rm W}$ and evaluate Eq.~\eqref{eq:weyl-canonical-surface} at the corresponding charge $q=\varrho q_c^{\rm W}$.

As in Gauss--Bonnet gravity, increasing the coupling lowers the transition temperature at fixed charge and moves the critical edge to smaller charge. At fixed coupling, the electric part of the first law also gives a charge slope for the Weyl-squared surface,
\begin{equation*}
 \left.\frac{\partial t_{\rm sl}^{\rm W}}{\partial q}\right|_{\lambda_{\rm W}}
 =\frac{2(\phi_l-\phi_s)}{s_l-s_s}.
\end{equation*}
Here $\phi_s,\phi_l$ and $s_s,s_l$ are evaluated on the coexisting Weyl-squared saddles. For positive charge and nonnegative coupling, the electrostatic potential in Eqs.~\eqref{eq:weyl-thermo-zero} and \eqref{eq:weyl-thermo-linear} decreases with radius to the order retained. The entropy increases with radius in the perturbative domain. The transition temperature thus decreases with charge as well. At the critical edge the two radii merge and the Wald entropy gap and latent heat vanish. A charge just below the Einstein critical value can therefore lie beyond the Weyl-squared critical edge at positive coupling, with consequences for the cross-CFT kink count in Sec.~\ref{sec:weyl-cuts}.

At zero charge, thermal AdS can be compared directly with the neutral black hole. In the Einstein limit they coexist at $x=1$ and $t_{\rm cyl}=3/(2\pi)$, with reduced entropy gap $4\pi$. The insertion in Eq.~\eqref{eq:weyl-insertion} vanishes on pure AdS and gives $w(1,0)=-72$ on the black hole. The temperature derivative of the black-hole-minus-AdS free energy is minus the entropy gap. Its zero therefore shifts to
\begin{equation*}
 t_{\rm HP}^{\rm W}
 =\frac{3}{2\pi}+\lambda_{\rm W}\frac{w(1,0)}{4\pi}
     +O(\lambda_{\rm W}^2)
 =\frac{3}{2\pi}-\frac{18}{\pi}\lambda_{\rm W}+O(\lambda_{\rm W}^2).
\end{equation*}
Here $t_{\rm HP}^{\rm W}$ is the reduced Weyl-squared Hawking--Page temperature.

\subsection{Coexistence at matched anomaly and current data}
\label{sec:matched-coexistence}

The canonical surfaces describe paths at fixed normalization within each family. We can also compare the transitions along paths with matched anomaly coefficients, current normalization and physical flux, as in Eqs.~\eqref{eq:anomaly-match} and \eqref{eq:charge-match}. Use the Weyl-squared coupling to parameterize both paths. As in Eq.~\eqref{eq:weyl-switch}, the first-order shift is the perturbation to the large-minus-small free-energy difference divided by the Einstein entropy gap. Let $t_{\rm sl}^{\rm GB,matched}$ denote the reduced cylinder transition temperature on the matched Gauss--Bonnet path. Evaluating Eq.~\eqref{eq:matched-free} at the two Einstein radii gives
\begin{equation}
 \left.\frac{d}{d\lambda_{\rm W}}
    \bigl(t_{\rm sl}^{\rm W}-t_{\rm sl}^{\rm GB,matched}\bigr)\right|_0
 =\frac{57 b^2}{4\Delta s_E}
   \left(x_s^{-8}-x_l^{-8}\right)>0 .
 \label{eq:matched-switch}
\end{equation}
This holds at fixed positive subcritical charge with distinct stable saddles. The residual interaction lowers the Weyl-squared free energy by a larger amount on the small saddle and raises the transition temperature relative to the matched Gauss--Bonnet theory at the same perturbative order. Starting from the common Einstein point at $q=0.05$, the derivatives along the two matched paths are approximately $-5.24848$ and $-5.60538$, respectively. The difference survives because a common vacuum energy cancels within each canonical comparison.

For the neutral Hawking--Page comparison the residual term in Eq.~\eqref{eq:matched-free} vanishes, and matching the Euler coefficients also matches the cylinder vacuum energies. The two neutral transition temperatures therefore agree through first order along the matched paths. The next section uses the charged canonical minima to locate coexistence across the theories in each family.

\section{Coexistence across the theory families}
\label{sec:phase-maps}

The canonical minima of Sec.~\ref{sec:fixed-theory} determine the coexistence source for a specified pair through Eq.~\eqref{eq:cylinder-boundary}. To extend this construction to a continuous interval, we test endpoint selection in the reference HR prescription with a linear source. Transforming the control together with the energy as in Appendix~\ref{app:scheme-comparison} preserves the full potential, hence its minimizing couplings and coexistence structure. We then compare temperature--source cuts for both families at the same benchmark charge.

\subsection{Gauss--Bonnet endpoint selection}
For Gauss--Bonnet gravity we work on $0\leq\lambda\leq\lambda_U$ at fixed $C,R$, nonzero $q$, proper bath temperature and source. The upper bound $\lambda_U\leq0.09$ follows from planar boundary microcausality \cite{Brigante2008Causality,BuchelMyers2009Causality}. Those references use the coefficient $\lambda_{\rm GB}L^2/2$ for the Gauss--Bonnet density, so $\lambda_{\rm GB}=\lambda$. Throughout, stability refers to radial thermodynamic stability of the spherical saddles.

Gauss's law makes the flux of smooth thermal AdS vanish in the absence of charged matter. The candidates at nonzero charge are therefore spherical black holes on the Einstein-connected branch. The charge convention of Sec.~\ref{sec:framework} fixes $q$ across the interval, and we plot $q>0$ since energy and temperature depend on $q^2$. The numerator of Eq.~\eqref{eq:temperature} has a unique positive zero that fixes the extremal radius on every slice. The denominator and clock factor are positive for $x>0$. Thus a radius with positive state temperature on any slice is admissible at both endpoints.

This common radius domain allows a pointwise comparison before radial minimization. We vary the coupling at fixed radius, cylinder bath temperature $t_{\rm cyl}$ and source $z_{\rm cyl}$. The state temperature can vary with coupling during this comparison. Once the endpoints are selected, radial minimization in each theory gives the thermal saddles whose state temperature matches the bath.

In units of $C/R$ the potential is $g_{\rm cyl}=R\mathcal G_{\rm cyl}/C$. Combining Eqs.~\eqref{eq:grand} and \eqref{eq:proper-time} gives
\begin{equation}
g_{\rm cyl}=3\beta(m+\lambda)+e_{\rm vac}^{\rm GB}(\lambda)
 -t_{\rm cyl}s_{\rm GB}-z_{\rm cyl}\lambda.
\label{eq:cylinder-map}
\end{equation}
At fixed radius, the entropy and source contributions are affine in $\lambda$. The energy terms determine the curvature
\begin{equation}
\partial_\lambda^2g_{\rm cyl}
=3\beta''(m+\lambda)-\frac{3}{16\eta\beta}<0,
\end{equation}
Primes here denote coupling derivatives at fixed radius and cylinder data. The vacuum and mass terms give $6\beta'+(e_{\rm vac}^{\rm GB})''=-3/(16\eta\beta)$ and $\beta''=-1/(\eta^3\beta)-1/(4\eta^2\beta^3)$. Strict concavity places the potential at every interior coupling above the smaller endpoint value at the same radius. Minimizing over the radius at either endpoint can only lower its potential, so two radial minimizations suffice for the global comparison.

\subsection{Weyl-squared endpoint selection}
\label{sec:weyl-selection}

To first order in $\lambda_{\rm W}$, the free energy along each continued Weyl-squared thermal branch is affine in the coupling. If the same branch gives both endpoint minima at coexistence, its shifted free energy is then constant along the interval at that order. To establish strict endpoint selection, we compute the second coupling derivative at the Einstein point for the action in Eq.~\eqref{eq:actions}, with independent higher-order Wilson coefficients set to zero.

At fixed bath temperature, flux and boundary geometry, the first coupling derivative of the reduced canonical free energy is the explicit Weyl-squared insertion on the deformed saddle in the same units. Differentiating the insertion requires the first-order metric and the displacement of the horizon at fixed bath temperature. This gives the curvature (Appendix~\ref{app:weyl-curvature})
\begin{equation}
 K_t\equiv\left.\partial_{\lambda_{\rm W}}^2 h_{\rm on}\right|_0
   =-\frac{6\mathcal P(y,u)}{7y(5u+2y-1)},\qquad
 y=x^2,\quad u=b/x^4,
 \label{eq:weyl-curvature}
\end{equation}
The numerator polynomial $\mathcal P(y,u)$ is positive for $y>0$ and $u\geq0$, as shown in the same appendix. The denominator has the sign of $\tau_0'$ since $5u+2y-1=2\pi y\,\tau_0'$. Thus $K_t$ is negative on each nondegenerate, radially stable Einstein branch. To isolate the radial contribution, let $K_x$ be the coupling curvature of the off-shell potential at fixed radius and bath temperature. Then
\begin{equation}
 K_t=K_x-\frac{\tau_1^2 s_0'}{\tau_0'} .
 \label{eq:radial-relaxation}
\end{equation}
At $q=0.05$ and the Einstein coexistence temperature, the small and large saddles have fixed-radius curvatures $K_x\simeq116.76$ and $0.2291$. Radial relaxation makes their curvatures $-293.99$ and $-1912.37$, respectively. In this example it reverses the sign on both branches.

At a fixed subcritical charge, the two competing Einstein saddles are nondegenerate near the small--large coexistence temperature. Their negative curvatures persist by smooth continuation on a sufficiently small positive coupling interval. Adding the linear source term leaves each branch concave. Taking the smaller of their two potentials at each coupling then gives the endpoint bound of Sec.~\ref{sec:identities}. This establishes local endpoint selection among the competing spherical Einstein-connected saddles. A quantitative interval width would require control of higher-order terms. Independent interactions at the same perturbative order can alter Eq.~\eqref{eq:weyl-curvature}, as illustrated in Appendix~\ref{app:weyl-curvature}.

\subsection{Gauss--Bonnet coexistence cuts}

For the Gauss--Bonnet interval, the endpoint proof reduces the phase diagram to the canonical minima of the two endpoint theories. Equation~\eqref{eq:canonical-coex-slope} implies that the upper theory switches first whenever both theories have small--large transitions.

The critical charges $q_{{\rm c},U}=q_c(\lambda_U)$ and $q_{{\rm c},E}=q_c(0)$ determine how many switches occur, as summarized in Table~\ref{tab:kink-count}. For $\lambda_U\geq1/36$ the upper endpoint has a unique radial minimum at each positive bath temperature and nonzero charge, since the polynomial in Eq.~\eqref{eq:turning-polynomial} is positive there. We use $q_{{\rm c},U}=0$ for this case in the table.

\begin{table}[tbp]
\caption{Kinks on the cross-CFT coexistence curve when the upper theory has the smaller critical charge and, when both switches occur, the lower transition temperature. The table applies to the Gauss--Bonnet intervals studied and describes the local Weyl-squared pattern near the critical charges as discussed in Sec.~\ref{sec:weyl-cuts}.}
\label{tab:kink-count}
\centering
\begin{tabular}{ll}
\hline\hline
Charge range & Switches in order of increasing temperature\\
\hline
$0<q<q_{{\rm c},U}$ & Two (upper, then Einstein)\\
$q_{{\rm c},U}<q<q_{{\rm c},E}$ & One (Einstein)\\
$q>q_{{\rm c},E}$ & Zero\\
\hline\hline
\end{tabular}
\end{table}

At $q=q_{{\rm c},j}>0$, the two radial minima of endpoint $j$ merge and their entropy gap vanishes. Its triple line then ends at a critical point on the cross-CFT boundary. The corresponding kink closes, while the other theory still contributes a distinct saddle. At $q=q_{{\rm c},U}>0$ the Einstein kink remains. At $q=q_{{\rm c},E}$ the Einstein kink has closed as well. The discontinuity in the preferred coupling remains $\lambda_U$ at either critical endpoint. The Einstein parameterization in Appendix~\ref{app:repro} gives
\begin{equation}
\frac{\Delta s_E}{\lambda_U}
=\frac{4\sqrt{2}\,5^{1/4}\pi}{\lambda_U}\sqrt{q_{{\rm c},E}-q}
+O(q_{{\rm c},E}-q).
\end{equation}

\begin{figure}[tbp]
\centering
\includegraphics[width=.96\linewidth]{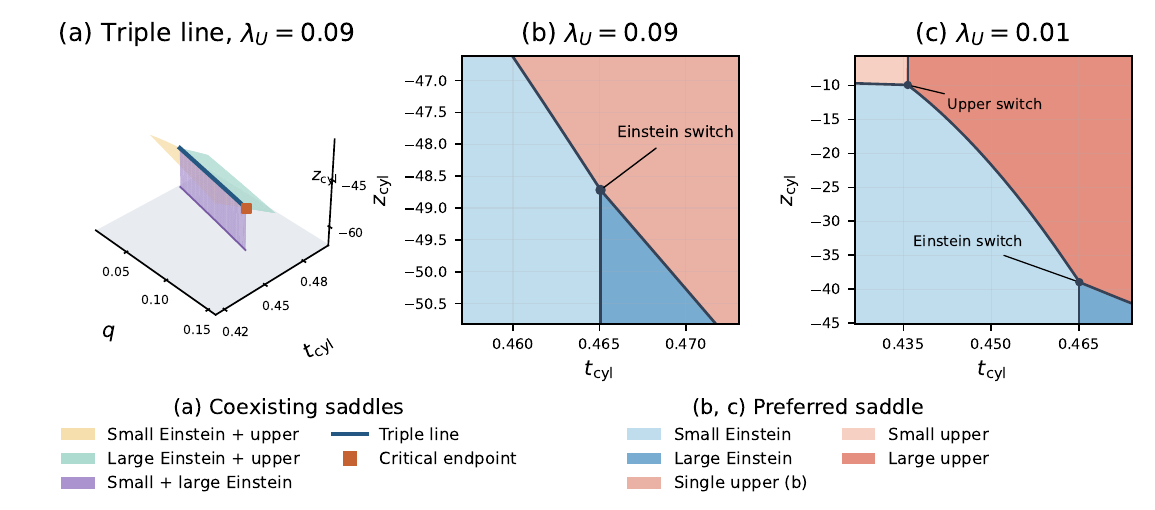}
\caption{Gauss--Bonnet coexistence in the zero-type-D holographic renormalization scheme, from Eqs.~\eqref{eq:fig-gb-minima}--\eqref{eq:fig-triple-rays} and \eqref{eq:cylinder-boundary}. The axes show reduced charge $q$, cylinder bath temperature $t_{\rm cyl}$ and source $z_{\rm cyl}$. In (a), $\lambda_U=0.09$. The gold and green sheets pair the small and large Einstein saddles, respectively, with the upper-endpoint saddle. The purple sheet is Einstein small--large coexistence, shown down to the lower plotting cutoff. The blue triple line ends at the square, where the two Einstein saddles merge. Panels (b) and (c) fix $q=0.05$ at $\lambda_U=0.09$ and $0.01$, respectively. The shared color legend identifies the preferred saddle, including the single upper saddle in (b). Dark curves mark cross-CFT coexistence, vertical rays mark small--large coexistence within the preferred theory and dots mark triple points.}
\label{fig:cylinder-triple}
\end{figure}

To construct Fig.~\ref{fig:cylinder-triple}, we solve the Gauss--Bonnet thermal equation \eqref{eq:radial-polynomial} for its positive roots at each $q,t_{\rm cyl}$ in each endpoint theory. The canonical minimum is
\begin{equation}
 H_j=\min_{x\, :\, \tau_{\rm cyl}=t_{\rm cyl}}
 \bigl[3\beta(\lambda_j)(m+\lambda_j)+e_{\rm vac}^{\rm GB}(\lambda_j)
       -t_{\rm cyl}s_{\rm GB}(x,\lambda_j)\bigr],
 \qquad j\in\{E,U\}.
 \label{eq:fig-gb-minima}
\end{equation}
Equation~\eqref{eq:cylinder-boundary} gives the cross-CFT boundary. The upper-minus-Einstein potential difference is $\lambda_U(z_{\rm coex}-z_{\rm cyl})$, so the color on each side identifies the preferred theory and the shade identifies its minimizing radial branch. Let $t_{{\rm sl},j}(q)$ be the cylinder temperature of a small--large transition in endpoint $j$. Its coexistence ray lies at $t_{\rm cyl}=t_{{\rm sl},j}$ at fixed charge. The triple-point source $z_{{\rm tr},j}$ and the range of sources along each ray are
\begin{equation}
 z_{{\rm tr},j}=z_{\rm coex}(t_{{\rm sl},j},q),\qquad z_{\rm cyl}\le z_{{\rm tr},E}\ \ (j=E),\qquad z_{\rm cyl}\ge z_{{\rm tr},U}\ \ (j=U).
 \label{eq:fig-triple-rays}
\end{equation}
The inequalities keep each ray in the region where its theory has the lowest potential. The same construction will be used for the Weyl-squared cuts.

The two cross-CFT sheets in panel (a) are $(q,t_{\rm cyl},z_{\rm coex}(t_{\rm cyl},q))$ on the two sides of $t_{\rm cyl}=t_{{\rm sl},E}(q)$. Their intersection is the triple line, parameterized by $q(k)$ and $t_{{\rm sl},E}(k)$ from Eq.~\eqref{eq:triple-param}. Sweeping the Einstein ray of Eq.~\eqref{eq:fig-triple-rays} over charge gives the purple sheet. The square marks the $k\to1$ limit at $q=q_{{\rm c},E}$ and $t_{\rm cyl}=t_{{\rm c},E}$. There the two Einstein radii and their entropies coincide, giving the two cross-CFT sheets the same temperature slope. The upper theory remains distinct at that endpoint.

The slice $\lambda_U=0.01$ marked in Fig.~\ref{fig:charged-surface} illustrates two switches. Its critical charge is $q_{{\rm c},U}\simeq0.10591105$, so the benchmark $q=0.05$ lies below both critical charges. The two canonical transitions occur at distinct temperatures and give the triple points
\begin{equation}
 (t_{\rm cyl},z_{\rm cyl})_{\rm U}\simeq(0.43570554,-9.94243806),
 \qquad (t_{\rm cyl},z_{\rm cyl})_E\simeq(0.46506518,-38.95187444).
\end{equation}
The three reduced potentials coincide at each point, with values $0.89348866$ and $0.89049865$, respectively. The one-sided slopes give jumps of $-628.1182$ and $+954.0350$, in agreement with Eq.~\eqref{eq:triple-crease}. The upper critical endpoint lies at $t_{\rm cyl}\simeq0.42238980$ and $z_{\rm cyl}\simeq-14.58675760$.

At the same charge with $\lambda_U=0.09$, the upper theory has a single radial minimum. The Einstein switch gives the triple point $(t_{\rm cyl},z_{\rm cyl})\simeq(0.46506518,-48.71743621)$ with common reduced potential $0.89049865$ and slope jump approximately $106.0039$. On this interval the Einstein critical endpoint is at $t_{\rm cyl}=t_{{\rm c},E}$ and $z_{\rm cyl}\simeq-41.47156257$.

The different patterns in panels (b) and (c) follow from the canonical phase structure of the upper theory. In (b) the single kink comes from the Einstein entropy gap. In (c) the upper theory already prefers a large black hole while Einstein still prefers a small one. The upper switch increases $s_U$ and lowers the boundary slope. The later Einstein switch increases $s_E$ and raises the slope, as in Eq.~\eqref{eq:triple-crease}.

\subsection{Weyl-squared coexistence cuts}
\label{sec:weyl-cuts}

For sufficiently small positive $\lambda_{\rm W,U}$, Eq.~\eqref{eq:weyl-critical-edge} places the upper critical charge $q_{{\rm c},U}=q_c^{\rm W}(\lambda_{\rm W,U})$ below $q_{{\rm c},E}$. Where both canonical transitions exist, Eq.~\eqref{eq:weyl-switch} gives the upper switch first at fixed subcritical charge. This ordering extends to the local critical region because the fixed-charge shift has a finite negative limit as $k\to1$ in Eq.~\eqref{eq:weyl-canonical-surface}. The endpoint pair therefore follows Table~\ref{tab:kink-count} near the critical charges, with the upper threshold known to the order in Eq.~\eqref{eq:weyl-critical-edge}.

Varying the charge traces two local triple lines through Eq.~\eqref{eq:fig-triple-rays}, one for each canonical switch. Each line ends on the cross-CFT boundary at the corresponding canonical critical edge, where its two same-theory saddles merge. Appendix~\ref{app:weyl-canonical} gives the square-root closure of the entropy gap there. By Eq.~\eqref{eq:triple-crease} the slope jump closes with the same power, while sweeping the source through coexistence still switches between the two endpoint couplings. The Einstein entropy gap is the same in both families because their lower endpoints are identical.

\begin{figure}[tbp]
\centering
\includegraphics[width=\linewidth]{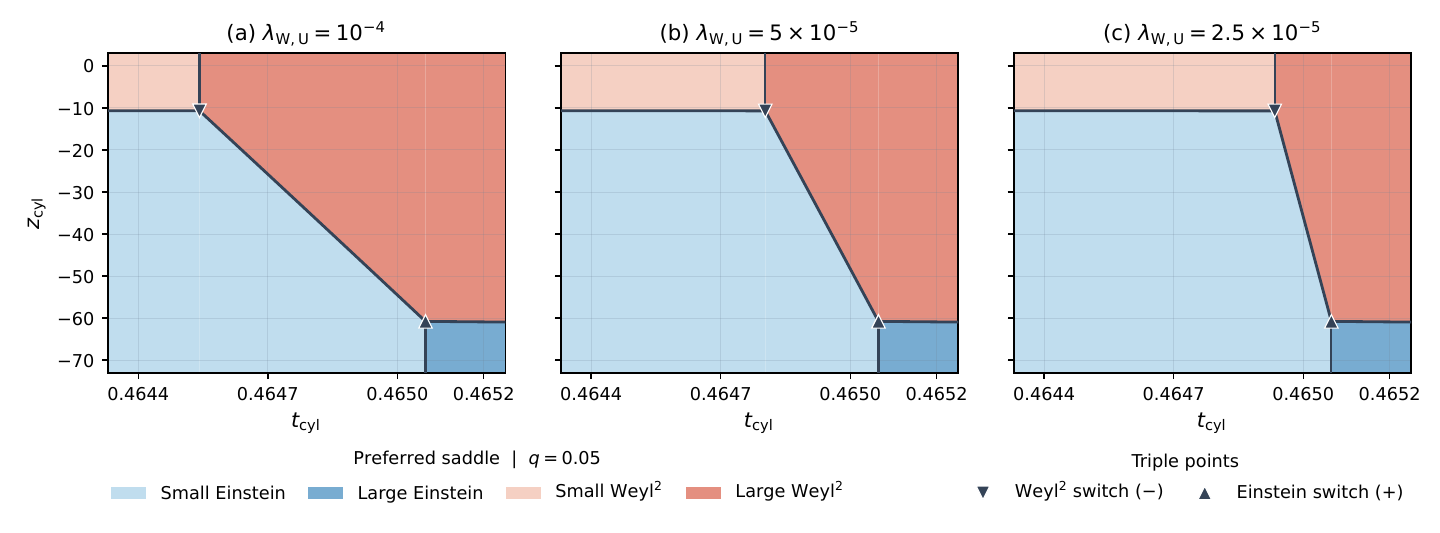}
\caption{Cross-CFT coexistence of Einstein and Weyl-squared endpoint theories at reduced charge $q=0.05$. Panels (a), (b) and (c) use upper Weyl-squared couplings $\lambda_{\rm W,U}$ of $10^{-4}$, $5\times10^{-5}$ and $2.5\times10^{-5}$ on common axes for the reduced cylinder bath temperature $t_{\rm cyl}$ and source $z_{\rm cyl}$. The dark curve is computed from the first-order free energies in Eq.~\eqref{eq:weyl-onshell} and the equality in Eq.~\eqref{eq:cylinder-boundary}. Blue and warm colors identify the preferred Einstein and Weyl-squared saddles, with light and dark shades for small and large black holes. Downward and upward triangles mark the Weyl-squared and Einstein switches, respectively. The signs in the legend give the change in the coexistence-curve slope as temperature increases, as in Eq.~\eqref{eq:triple-crease}. The upward Weyl-squared ray and downward Einstein ray follow Eq.~\eqref{eq:fig-triple-rays}.}
\label{fig:weyl-cuts}
\end{figure}

To construct the fixed-charge cuts in Fig.~\ref{fig:weyl-cuts}, we solve $\tau_0(x_{0,\gamma},b)=t_{\rm cyl}$ with $\tau_0'(x_{0,\gamma},b)>0$ for the reference radii of the small and large branches, $\gamma\in\{s,l\}$. The fixed-charge free energies in Eq.~\eqref{eq:weyl-onshell}, evaluated at these radii, give the reduced free energy $h_{j,\gamma}$ for each branch in endpoint theory $j$. The couplings are $\lambda_E=0$ and $\lambda_U=\lambda_{\rm W,U}$, where $\lambda_{\rm W,U}$ is the upper Weyl-squared coupling. Taking $H_j=\min(h_{j,s},h_{j,l})$ in Eq.~\eqref{eq:cylinder-boundary} gives the cross-CFT coexistence curve. Solving $h_{j,l}-h_{j,s}=0$ in each theory gives the switch temperatures. The triple points and rays follow from Eq.~\eqref{eq:fig-triple-rays}. Because Eq.~\eqref{eq:cylinder-boundary} divides the free-energy difference by $\lambda_{\rm W,U}$, terms omitted at second order in the free energy enter the source at order $\lambda_{\rm W,U}$.

In the first-order calculation at the largest coupling width, the Weyl-squared triple point has reduced cylinder temperature $0.46454187$ and source $-10.69565$. The corresponding Einstein values are $0.46506518$ and $-60.77836$. The three reduced saddle potentials coincide at each triple point, at $0.89055379$ and $0.89049865$, respectively. Equation~\eqref{eq:triple-crease} gives slope jumps of approximately $-95047.04$ and $+95403.50$ from the two Wald entropy gaps, in agreement with one-sided derivatives of the free-energy difference. The central segment pairs a small Einstein black hole with a large Weyl-squared black hole. Its entropy difference is larger than on the outer segments, where both saddles lie on the same type of radial branch.

At this charge Eq.~\eqref{eq:weyl-switch} gives $dt_{\rm sl}^{\rm W}/d\lambda_{\rm W}|_0\simeq-5.24848014$. Halving the coupling successively from $10^{-4}$ gives triple-point temperature separations $5.23309\times10^{-4}$, $2.62041\times10^{-4}$ and $1.31116\times10^{-4}$. Minimization at fixed entropy coordinate gives the same first-order shift. The difference between the two implementations falls by about a factor of four at each halving.

The limit of small coupling explains why the two triple points approach the same temperature but retain distinct sources. Let $w_s,w_l$ be the insertions on the two Einstein saddles at $t_{{\rm sl},E}$. At the Weyl-squared triple point, both Weyl-squared branches have the same free energy and the Einstein minimum is on the small branch. Thus the numerator of Eq.~\eqref{eq:cylinder-boundary} can be evaluated on the two small saddles. At the Einstein triple point the Weyl-squared theory has already switched, so both large saddles can be used. In each case the zeroth-order free energies cancel. Writing $\Delta t_{\rm sl}=t_{{\rm sl},E}-t_{{\rm sl},U}$ and $\lambda_U=\lambda_{\rm W,U}$, the limit $\lambda_U\to0$ gives
\begin{equation}
 \Delta t_{\rm sl}=\lambda_U\frac{w_s-w_l}{\Delta s_E}+O(\lambda_U^2),\qquad
 z_{{\rm tr},U}\to w_s,\qquad z_{{\rm tr},E}\to w_l.
 \label{eq:fig-weyl-triple-limit}
\end{equation}
The source separation tends to $w_s-w_l\simeq50.07234$, although the temperature separation decreases linearly at leading order. In that shrinking interval the two preferred saddles approach different Einstein branches, so their entropy gap stays finite. Equation~\eqref{eq:cylinder-clapeyron} therefore gives a central slope asymptotic to $-\Delta s_E/\lambda_{\rm W,U}$. The plots compare specified endpoint pairs. The separate concavity argument in Sec.~\ref{sec:weyl-selection} establishes when these endpoints minimize over a continuous interval.

\section{Discussion}
\label{sec:discussion}

Canonical transitions within each CFT leave kinks on the cross-CFT boundary. When both endpoints switch in succession, the upper theory switches first and gives a negative slope jump. The Einstein switch gives a positive one. The two canonical critical charges determine whether the endpoint coexistence curve has two, one or zero kinks, with the Weyl-squared count established locally near the critical edges. Each jump vanishes as the corresponding black holes merge. Sweeping the source across coexistence for the endpoint pair still gives a coupling jump equal to the interval width.

The kink structure describes competition between the endpoints once they are selected. The concavity that selects them depends on the curvature interaction. Gauss--Bonnet gravity has negative coupling curvature already at fixed radius. In the Weyl-squared calculation the fixed-radius curvatures are positive at the benchmark saddles. Allowing the horizon radius to respond at fixed temperature makes the curvatures of the thermal branches negative there. The relaxation term in Eq.~\eqref{eq:radial-relaxation} is proportional to the inverse radial stiffness at fixed mixed derivative. The curvature test is local to the nondegenerate Einstein branches and uses the second-order continuation of the action with the Weyl-squared term as its sole correction. The Weyl-squared coexistence curves for specified endpoint pairs use only first-order free energies.

The cylinder vacuum energy sets the conversion between the numerical sources of two energy prescriptions. At fixed normalization the Euler anomaly varies along the Gauss--Bonnet path, so the coexistence curve is translated in the source direction. Along the Weyl-squared path the Euler coefficient stays constant at the order studied, so the two Weyl-squared curves in Fig.~\ref{fig:schemes} coincide at the same numerical source. The Weyl-squared curves illustrate this covariance on the perturbative Einstein-connected branch, with a linear reference source and the matched controls of Eq.~\eqref{eq:scheme-compensation}.

The anomaly coefficients also provide a way to compare the two families at matched boundary data. Near the Einstein point they give local coordinates when both $C$ and the curvature coupling vary. The matched families have the same first-order neutral Hawking--Page temperature. At finite charge, matching the anomalies and the current two-point normalization leaves the term in Eq.~\eqref{eq:matched-free}. It is quartic in the flux at fixed reference radius and larger in magnitude on the small black hole. Consequently the Weyl-squared canonical transition occurs at a higher temperature than the matched Gauss--Bonnet transition at the same perturbative order, as quantified by Eq.~\eqref{eq:matched-switch}. The difference of the two saddle free energies removes the vacuum contribution within each theory. The anomaly trajectories and the charged canonical response therefore probe different information about the curvature deformation.

These conclusions concern classical spherical saddles at a prescribed source. To define a microscopic ensemble across theories, one must specify the sectors, the allowed couplings, how their charges are identified and how the sectors are weighted. Vacua or flux sectors can provide settings for such a construction \cite{KlebanovWitten1999Vacua,FreedmanGubserPilchWarner1999,BoussoPolchinski2000}. Under the support and subexponential-weight conditions in Appendix~\ref{app:measure}, the classical minimum gives the leading free energy. At coexistence the theory measure and the saddle prefactors determine relative probabilities. Exponential differences in sector weights can also change the leading phase competition.

\acknowledgments
Limin Zeng would like to express his deepest and foremost thanks to his partner Lu Lu.

\paragraph{Data availability.}
The accompanying software, \emph{Thermodynamics within and across holographic CFTs}, version 1.0.0, provides the code for all six figures, the benchmark data and the symbolic and numerical checks for both families \cite{ZengFigureCode}.

\appendix
\section{Radial minimization and canonical benchmarks}
\label{sec:checks}
\label{app:repro}
The phase diagrams use canonical minima evaluated at the bath temperature. For Gauss--Bonnet gravity, imposing radial stationarity in the cylinder clock gives
\begin{equation}
\beta(12x^6+6x^4-2q^2)-12\pi t_{\rm cyl}x^3(x^2+2\lambda)=0.
\label{eq:radial-polynomial}
\end{equation}
At positive bath temperature, every positive real root lies above extremality. The radial derivative is negative just above the extremal radius and the potential grows at large radius. The minimum must therefore occur at a root and is found by comparing the potential at all positive real roots.

At the Einstein endpoint the background-subtracted on-shell canonical free energy in units of $C/R$ is $-x^4+x^2+5q^2/(3x^2)$. Let $x_s$ and $x_l$ be the small and large radii in units of $L$. Set $k=x_s/x_l$ and $v=q^2/x_l^4$, with $0<k<1$. The first equation imposes equal temperatures and the second imposes equal on-shell free energies,
\begin{gather*}
6(1-k)x_l^2+3(1-k^{-1})-(1-k^{-5})v=0,\\
-(1-k^4)x_l^2+1-k^2+\tfrac53(1-k^{-2})v=0 .
\end{gather*}
Solving the linear system fixes the radii and charge on the Einstein coexistence curve. Substituting into the temperature gives $t_{{\rm sl},E}$, the reduced coexistence temperature where the two clocks coincide,
\begin{gather}
N_k=k^2+3k+1,\qquad D_k=k^4+3k^3+7k^2+3k+1,\notag\\
x_l^2=N_k/D_k,\qquad x_s=kx_l,\notag\\
q=\sqrt{3}\,k^2N_k/D_k^{3/2},\qquad
t_{{\rm sl},E}=\frac{3(1+k)^3}{2\pi\sqrt{N_kD_k}}.
\label{eq:triple-param}
\end{gather}
As $k$ increases from zero to one, the charge increases to $q_{{\rm c},E}$. In the interior the two radii in Eq.~\eqref{eq:triple-param} are minima separated by a maximum. At $\lambda=0$ Descartes' rule of signs bounds the number of positive roots of Eq.~\eqref{eq:radial-polynomial} by three. The degenerate minima therefore give $H_E$. With $t_{\rm cyl}=t_{{\rm sl},E}$ and the same charge, the upper-endpoint minimum follows from the Gauss--Bonnet radial equation or the Weyl-squared free energies in Eq.~\eqref{eq:weyl-onshell}.

\subsection{Canonical coexistence and critical points}
For Gauss--Bonnet gravity let $y=x^2$ and $\mathcal N(y)=6y^3+3y^2-q^2$. Differentiating Eq.~\eqref{eq:temperature} with respect to $y$ gives $2y(y+2\lambda)\partial_y\mathcal N-(5y+6\lambda)\mathcal N$ after positive factors are removed. Since $dy/dx>0$, the sign of $d\tau/dx$ is the sign of
\begin{equation}
\mathcal D(y)=3y^2[2y^2+(12\lambda-1)y+2\lambda]+q^2(5y+6\lambda).
\label{eq:turning-polynomial}
\end{equation}
For $1/36\leq\lambda\leq0.09$ and $q>0$, the polynomial is positive, giving a single spherical saddle at each positive bath temperature. For $0<\lambda<1/36$, let $y_-<y_+$ be the positive roots of the quadratic in brackets. A turning point lies between these roots and satisfies
\begin{equation}
q^2=W_{\rm sp}(y)\equiv\frac{6y^2(y-y_-)(y_+-y)}{5y+6\lambda}.
\label{eq:spinodal-charge}
\end{equation}
On this interval $\log W_{\rm sp}$ is strictly concave. The positive contribution of the denominator to its second derivative is less than $1/y^2$. The numerator contributes $-2/y^2-1/(y-y_-)^2-1/(y_+-y)^2$. Thus $W_{\rm sp}$ has a single maximum, which fixes $q_c^2$. At a turning point the temperature is positive because the derivative condition gives
\begin{equation}
(5y+6\lambda)\mathcal N=2y(y+2\lambda)\partial_y\mathcal N .
\end{equation}
Both turning points lie in the positive-temperature domain for $0<q<q_c$. They merge at $q=q_c$, and for $q>q_c$ the temperature is monotonic.

The radial derivative $(\tau-t)\partial_x(S/C)$ gives two minima and an intervening maximum between the turning temperatures. At the lower turning temperature the potential rises from the small minimum to the merged middle and large roots, so the large-minus-small difference is positive. At the upper turning temperature the potential falls from the merged small and middle roots to the large minimum, giving a negative difference. The large saddle has higher entropy, so the temperature derivative of the difference is negative throughout the interval. There is one crossing. The transition is first order for subcritical charge and its entropy jump vanishes when the two radii merge at $q_c$.

For Eq.~\eqref{eq:canonical-coex}, bracket the coexistence temperature by the two turning temperatures. At a trial temperature, solve $\tau=t$ for the smallest and largest positive radii and vary $t$ until their free-energy difference vanishes. The critical edge follows by maximizing $W_{\rm sp}$ between $y_-$ and $y_+$ and evaluating $\tau$ at the merged radius, as in Eq.~\eqref{eq:canonical-critical-edge}.

\subsection{Fixed-charge action and the theory measure}
\label{app:measure}

Let $\sigma$ label the energy prescription and $\beta_{\rm th}=1/\Theta_{\rm cyl}$ be the inverse bath temperature. The Maxwell boundary Legendre term converts the fixed-potential action into the fixed-charge action \cite{HawkingRoss1995,ChamblinFluctuations1999}. At a thermal saddle it gives
\begin{equation}
 I_\sigma^{\mathcal Q}=\beta_{\rm th}E_\sigma-S.
 \label{eq:fixed-charge-action}
\end{equation}
This action sets the saddle weight in the single-theory partition function $\mathcal Z_{\sigma,\lambda}^{\mathcal Q}$. To sum over theories, specify the set $\mathfrak T$ of allowed couplings and a positive finite measure $d\nu_C(\lambda)$ on it. The measure may depend on the common normalization $C$. The external control energy $U_\sigma(\lambda)$ equals $\xi_{\rm cyl}\lambda$ in the reference prescription of Eq.~\eqref{eq:grand}. The resulting partition function is
\begin{equation}
 \mathcal Z_{{\rm ext},\sigma}
  =\int_{\mathfrak T}d\nu_C(\lambda)\,
   e^{\beta_{\rm th}U_\sigma(\lambda)}\mathcal Z_{\sigma,\lambda}^{\mathcal Q}.
 \label{eq:theory-mixture}
\end{equation}
For discrete sectors the integral is a sum with the assigned weights. Conditioning on $\lambda$ recovers the canonical Gibbs state of that theory. A change of coordinate pushes the measure forward and composes the action and control with the inverse map. A density constant in $\lambda$ generally varies in the new coordinate. At fixed clock, the compensating change of $U_\sigma$ in Eq.~\eqref{eq:scheme-compensation} preserves the integrand for the same measure.

Let $H_\sigma(\lambda)$ be the minimum reduced canonical free energy and $g_\sigma(\lambda)=H_\sigma(\lambda)-RU_\sigma(\lambda)/C$. Take large $C$ at fixed $R$, reduced charge $q$, bath temperature $t_{\rm cyl}$ and normalized control $RU_\sigma/C$. For a saddle approximation uniform in $\lambda$, the exponent in Eq.~\eqref{eq:theory-mixture} is $-Cg_\sigma/t_{\rm cyl}+o(C)$ with a sublinear remainder. Assume that the total mass of the measure grows at most subexponentially. Also assume that $g_\sigma$ is continuous at a global minimum and the logarithm of the measure mass of each fixed neighborhood of that minimum has a sublinear lower bound. These conditions give
\begin{equation}
 -\lim_{C\to\infty}\frac{t_{\rm cyl}}{C}\log\mathcal Z_{{\rm ext},\sigma}
     =\min_{\lambda\in\mathfrak T}g_\sigma(\lambda).
 \label{eq:theory-laplace}
\end{equation}
For an upper bound on the partition function, multiply $\exp[-C\min g_\sigma/t_{\rm cyl}+o(C)]$ by the total weight of the measure. For a lower bound, restrict the integral to a fixed neighborhood where $g_\sigma$ exceeds its minimum by at most $\epsilon>0$. The measure factors drop out at leading order under the stated conditions, and one can then let $\epsilon$ tend to zero. A continuous density whose support reaches an endpoint can therefore yield endpoint dominance even if the point itself has zero measure. At endpoint coexistence, both endpoints contribute at the same exponential order if $g_\sigma$ is continuous at each endpoint and every fixed neighborhood of each endpoint satisfies the mass condition.

The measure and the factors multiplying the saddle exponentials determine the relative probabilities when the classical potentials agree. A discrete endpoint mass and an integral over a continuous neighborhood can give different powers of $C$. For example, a smooth nonzero density at an isolated endpoint with nonzero one-sided potential slope gives a neighborhood factor proportional to $C^{-1}$. Let $A_E$ and $A_U$ denote the combined measure and saddle prefactors for two isolated competing endpoints. In the reference HR prescription with a linear control, $g_U-g_E=-\lambda_U(z_{\rm cyl}-z_{\rm coex})$, where $g_j=g_{\rm HR}(\lambda_j)$. Equating the weights $A_j\exp[-Cg_j/t_{\rm cyl}]$ gives the leading displacement of the equal-weight source,
\begin{equation}
 z_{\rm equal}-z_{\rm coex}
   \simeq -\frac{t_{\rm cyl}}{C\lambda_U}\log\frac{A_U}{A_E}.
 \label{eq:measure-rounding}
\end{equation}
Evaluate the prefactors at the classical crossing if their relative variation over the displacement is negligible. At a triple point, each endpoint contributes the sum over its degenerate saddles with their own prefactors. The measure and the prefactors then determine their relative probabilities.

Exponential weights can also affect the leading phase competition. If the density relative to a fixed reference measure grows as $\exp[C\,\chi_{\mathfrak T}(\lambda)+o(C)]$, then $\chi_{\mathfrak T}$ is the logarithmic weight per unit $C$. Under the same saddle and support conditions for the reference measure, one minimizes $g_\sigma-t_{\rm cyl}\chi_{\mathfrak T}$ on the supported set. Removing a neighborhood of a minimizing endpoint from the support can also change the result. Equation~\eqref{eq:theory-mixture} thus gives a statistical interpretation of the classical phase diagrams once the weights and support are specified.

\section{Gauss--Bonnet cylinder energy and scheme comparison}
\label{app:stress}
\subsection{Cylinder energy from the stress tensor}
The energy in Eq.~\eqref{eq:proper-time} comes from the time projection of the quasilocal stress tensor. We use the Dirichlet momentum and intrinsic counterterms constructed by Anastasiou et al.\ \cite{Anastasiou2026}, with $\ell=L_{\rm eff}$. The bulk solution has the asymptotic form
\begin{equation}
f_{\rm GB}(r)=1+r^2/\ell^2-m_{\rm bulk}/(\eta r^2)+O(r^{-4}).
\end{equation}
The explicit electric term enters at order $r^{-4}$. Let $\gamma_{ij}$ be the induced metric at fixed $r$. With an outward normal, the extrinsic curvature $K^i{}_j$ has the following components. Here $t$ labels the bulk time direction, $a,b$ are sphere indices and $f_{\rm GB}'=df_{\rm GB}/dr$.
\begin{equation*}
K^t{}_t=\frac{f_{\rm GB}'}{2\sqrt{f_{\rm GB}}},\qquad
K^a{}_b=K_\Omega\delta^a_b,\qquad K_\Omega=\frac{\sqrt{f_{\rm GB}}}{r}.
\end{equation*}
The Brown--York metric variation \cite{BrownYork1993} gives the cutoff stress tensor $T_{ij}$, including the Myers term \cite{Myers1987} and intrinsic counterterms. To obtain the energy density $\rho_{\rm cut}=T_{ij}v^iv^j$, use the unit time vector $v^i$ with temporal component $1/\sqrt{f_{\rm GB}}$ and zero spatial components. The Einstein term in $8\pi G_5\rho_{\rm cut}$ is $K^t{}_t-K=-3K_\Omega$, with $K=K^i{}_i$. Projecting the Gauss--Bonnet part of their Dirichlet momentum (Eq.~(3.15)) onto the time direction gives the term below. The coefficient of the Gauss--Bonnet density in Eq.~\eqref{eq:actions} is $\alpha/2$, and $\mathcal S^a{}_b$ is the intrinsic Schouten tensor.
\begin{equation*}
-\alpha\delta^{abc t}_{def t}K^d{}_a
\left(2\mathcal S^e{}_b\delta^f_c-\tfrac13 K^e{}_bK^f{}_c\right).
\end{equation*}
The antisymmetric delta restricts the contraction to the sphere, where $\mathcal S^a{}_b=\delta^a_b/(2r^2)$. This gives $-6\alpha K_\Omega/r^2+2\alpha K_\Omega^3$. The intrinsic counterterm density in their convention (Eq.~(3.12)) is $c_{\rm vol}+c_{\rm curv}\mathcal R[\gamma]$. The volume and scalar-curvature coefficients follow from their Eq.~(3.13),
\begin{equation*}
c_{\rm vol}=3/\ell-2\alpha/\ell^3,\qquad
c_{\rm curv}=\ell/4+\alpha/(2\ell).
\end{equation*}
Since $\mathcal R[\gamma]=6/r^2$ and the temporal Ricci component vanishes, the counterterm projection is $c_{\rm vol}+6c_{\rm curv}/r^2$. Adding it to the Einstein and Gauss--Bonnet projections yields
\begin{equation}
8\pi G_5\rho_{\rm cut}=-3K_\Omega+2\alpha K_\Omega^3+3/\ell-2\alpha/\ell^3+(3\ell/2+3\alpha/\ell-6\alpha K_\Omega)/r^2 .
\label{eq:cutoff-energy-density}
\end{equation}
The cutoff geometry is a product of time and a round three-sphere. Both its Weyl tensor and Euler density vanish, so the curvature-squared logarithmic variation gives zero. The zero-type-D prescription sets the finite scalar-curvature-squared counterterm to zero.

Using $\eta=1-2\alpha/\ell^2$, the required expansion is
\begin{equation}
K_\Omega=\frac1\ell+\frac{\ell}{2r^2}
-\frac{1}{r^4}\left(\frac{m_{\rm bulk}\ell}{2\eta}
+\frac{\ell^3}{8}\right)+O(r^{-6}).
\end{equation}
With this expansion the intrinsic counterterms cancel the terms of order $r^0$ and $r^{-2}$ in the cutoff energy density. The mass term in $r^4\rho_{\rm cut}$ comes from multiplying $-3\eta$ by $-m_{\rm bulk}\ell/(2\eta)$. The vacuum term is proportional to $3\ell^3(1-6\alpha/\ell^2)/8$. At this order the charge enters through the horizon mass. The boundary field strength vanishes and the Maxwell boundary term gives the fixed-charge Legendre transform in Eq.~\eqref{eq:fixed-charge-action}.

Rescaling to a boundary sphere of radius $R$ gives the unit-lapse energy per unit sphere volume as $\lim_{r\to\infty}r^4\rho_{\rm cut}/R$. Thus
\begin{equation}
E_{\rm cyl}^{\rm HR}=\frac{3}{16\pi G_5R}\left[m_{\rm bulk}\ell+\frac{\ell}{4}(\ell^2-6\alpha)\right]=\beta E_{R/L}^{\rm BS}+\frac{3a_{\rm n}}{4R}.
\label{eq:stress-energy}
\end{equation}
At zero coupling this reduces to the Einstein--AdS mass and cylinder Casimir energy.

\paragraph{Check from the bulk action and horizon data.}
To check the horizon quantities, use the static spherical ansatz of Eq.~\eqref{eq:weyl-metric} in units $L=1$, with radial function $f$, lapse $N$ and electric potential $\mathcal A_t$. Write $\mathcal A_\infty$ for its boundary value. Denote the signed sectional curvatures by $X_{t\rho},X_{t\Omega},X_{\rho\Omega},X_{\Omega\Omega}$. An $\Omega$ index indicates an angular direction. Primes denote $\rho$ derivatives, and the four curvatures are
\begin{equation*}
 X_{t\rho}=-\tfrac12 f''-\frac{3f'N'}{2N}-\frac{fN''}{N},\quad
 X_{t\Omega}=-\frac{f'}{2\rho}-\frac{fN'}{\rho N},\quad
 X_{\rho\Omega}=-\frac{f'}{2\rho},\quad
 X_{\Omega\Omega}=\frac{1-f}{\rho^2}.
\end{equation*}
The Gauss--Bonnet contraction is $24[X_{t\rho}X_{\Omega\Omega}+2X_{t\Omega}X_{\rho\Omega}+(X_{t\Omega}+X_{\rho\Omega})X_{\Omega\Omega}]$. The Maxwell first integral is $\rho^3\mathcal A_t'/N=P$. A Legendre transform eliminates the electric derivative at fixed flux. After dropping radial total derivatives, the action per unit sphere volume and unit time is proportional to the radial integral of
\begin{equation*}
 3N\partial_\rho\!\left[\rho^2(1-f)+\rho^4+\lambda_{\rm GB}(1-f)^2\right]
 -\frac{NP^2}{2\rho^3}.
\end{equation*}
Variation with respect to $f$ gives $N'=0$ on the Einstein branch. Varying the lapse and integrating gives the metric polynomial in Sec.~\ref{sec:families}. The electric term in the integrated mass function is $-b/\rho^2$ since $P^2=12b$. For $N=1$, differentiating the metric polynomial at the horizon gives $\tau=f'(x)/(4\pi)$, as in Eq.~\eqref{eq:temperature}. The Wald entropy is the horizon area density times $(1+\alpha\mathcal R_{S^3})/(4G_5)$. Using $\mathcal R_{S^3}=6/r_+^2$ gives Eq.~\eqref{eq:entropy}. The Maxwell potential drop is $\mathcal A_\infty-\mathcal A_t(x)=P/(2x^2)$, which becomes $\phi_{\rm GB}=\beta P/(2x^2)$ in the cylinder clock. Differentiating the stress-tensor energy at fixed coupling gives
\begin{equation*}
 \partial_x e_{\rm GB}=\beta(12x^3+6x-6b/x^3)=\tau_{\rm cyl}\partial_x s_{\rm GB},
 \qquad
 \partial_P e_{\rm GB}=\frac{\beta P}{2x^2},\qquad \partial_P s_{\rm GB}=0 .
\end{equation*}
The radial and electric derivatives thus agree with the horizon first law. The separate vacuum term in Eq.~\eqref{eq:stress-energy} enters when the coupling varies.

\subsection{Energy prescriptions for both families}
\label{app:scheme-comparison}
\label{sec:scheme-covariance}
Both background subtraction and local counterterms are available in higher-derivative AdS gravity \cite{NojiriOdintsovSurface2000}. The source translation was evaluated for the Gauss--Bonnet cylinder in our earlier work \cite{Zeng2026Revisiting} (Appendix C). At fixed clock the potential is $\mathcal G_{\sigma}[U_\sigma]=E_\sigma-\Theta_{\rm cyl}S-U_\sigma(\lambda)$. The reference convention uses $U_{\rm HR}=\xi_{\rm cyl}\lambda$. Under a reparametrization $\lambda\to\lambda^\prime$, the control is evaluated at the same theory as $U_\sigma(\lambda(\lambda^\prime))$. The conjugate becomes $\Xi_{\lambda^\prime}=\Xi_\lambda\,d\lambda/d\lambda^\prime$.

At fixed cylinder data, a finite curvature-squared counterterm changes $\sigma$ to $\sigma'$ and shifts the energy by $\Delta E=E_{\sigma'}-E_\sigma=(C/R)\Delta e(\lambda)$, with $\Delta e$ the reduced shift \cite{Assel2015,Anastasiou2026,Benjamin2024}. The factor $1/R$ follows from the volume scaling $R^3$ and the curvature-squared scaling $R^{-4}$. A state-independent shift changes the potential by
\begin{equation}
\mathcal G_{\sigma'}[U_{\sigma'}]-\mathcal G_{\sigma}[U_{\sigma}]
=\Delta E+U_{\sigma}-U_{\sigma'} .
\label{eq:scheme-compensation}
\end{equation}
The transformation $U_{\sigma'}(\lambda)=U_\sigma(\lambda)+\Delta E(\lambda)$ preserves the potential at every radius and coupling. To express it in reduced units, let $\mathcal U_\sigma=RU_\sigma/C$. Starting from $\mathcal U_\sigma=z_{\rm cyl}\lambda$, subtract the constant $\mathcal U_{\sigma'}(0)$ and write the result as $z_{\sigma'}(\lambda)\lambda$. Then
\begin{equation*}
 z_{\sigma'}(\lambda)=z_{\rm cyl}
      +\frac{\Delta e(\lambda)-\Delta e(0)}{\lambda},\qquad \lambda\ne0 .
\end{equation*}
The control after subtraction vanishes at $\lambda=0$ by construction. A constant source suffices throughout the interval precisely when $\Delta e(\lambda)$ is affine. For a nonaffine shift, minimizing $H_{\sigma'}-\mathcal U_{\sigma'}$ with the full compensated control gives the same globally preferred couplings as in the reference prescription. This preserves endpoint selection whenever it holds there. Restricting the transformed comparison to a constant linear source can change the interior potential and its minimum.

For the endpoint pair, a constant translation always suffices to match their potential difference. Requiring $\Delta E(\lambda_j)-\delta\xi\,\lambda_j$ to agree at $j=E,U$ gives
\begin{equation}
\delta\xi=\frac{\Delta E(\lambda_U)-\Delta E(0)}{\lambda_U}.
\end{equation}
This gives the reduced source map of Sec.~\ref{sec:framework} and leaves the radial minima unchanged.

\begin{figure}[tbp]
\centering
\includegraphics[width=.93\linewidth]{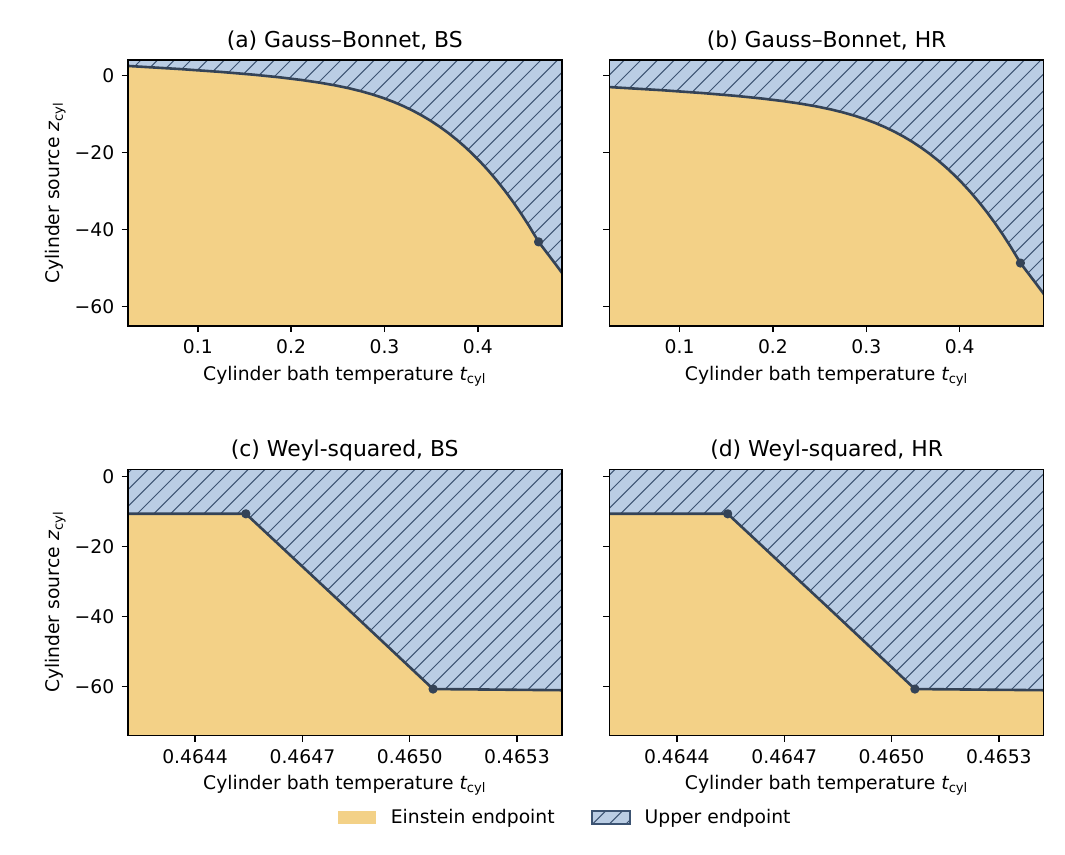}
\caption{Background subtraction (BS) and zero-type-D holographic renormalization (HR) at fixed $C,R$ and $q=0.05$ in the common cylinder clock. Equation~\eqref{eq:fig-scheme-curves} gives the boundary between the unhatched Einstein region and the hatched upper-endpoint region. In (a) and (b), Gauss--Bonnet with $\lambda_{\rm GB,U}=0.09$ translates the boundary downward by $5.48728$ from BS to HR. The dots mark the Einstein triple point in each prescription. In (c) and (d), the Weyl-squared boundaries and triple points coincide at first order for $\lambda_{\rm W,U}=10^{-4}$.}
\label{fig:schemes}
\end{figure}

For Fig.~\ref{fig:schemes}, $H_j^{\rm BS}$ is the minimum BS canonical free energy on the common cylinder in units of $C/R$. For Gauss--Bonnet gravity, compute it by omitting $e_{\rm vac}^{\rm GB}$ from Eq.~\eqref{eq:fig-gb-minima}. For Weyl-squared gravity, subtract $3/4$ from the endpoint free energies in Eq.~\eqref{eq:weyl-onshell} at the order shown. Define $e_{{\rm vac},j}=R(E_j^{\rm HR}-E_j^{\rm BS})/C$, and let $\delta_{\sigma,{\rm HR}}$ be one for HR and zero for BS. The boundary in either prescription is
\begin{equation}
 z_{\rm coex}^{\sigma}(t_{\rm cyl})=\frac{H_U^{\rm BS}-H_E^{\rm BS}+\delta_{\sigma,{\rm HR}}(e_{{\rm vac},U}-e_{{\rm vac},E})}{\lambda_U}.
 \label{eq:fig-scheme-curves}
\end{equation}
Each panel uses the numerical source in its own prescription. The upper theory is preferred in the hatched region $z_{\rm cyl}>z_{\rm coex}^\sigma$. The dots mark the canonical switch temperatures of Eq.~\eqref{eq:fig-triple-rays}, which are unchanged by the vacuum shifts.

On the Gauss--Bonnet cylinder, changing from BS to the zero-type-D scheme adds $\Delta E=Ce_{\rm vac}^{\rm GB}/R$, with $e_{\rm vac}^{\rm GB}$ from Eq.~\eqref{eq:proper-time}. The corresponding source translation is
\begin{equation}
 \delta z_{\rm cyl}=\frac{e_{\rm vac}^{\rm GB}(\lambda_U)-e_{\rm vac}^{\rm GB}(0)}{\lambda_U}.
 \label{eq:scheme-source-shift}
\end{equation}
The BS potential on the cylinder is also strictly concave at fixed radius, since $3\beta''(m+\lambda)+6\beta'<0$. Thus both prescriptions select from the same two endpoints. In the top row of Fig.~\ref{fig:schemes}, $\lambda_U=0.09$ and $q=0.05$. The BS triple source is approximately $-43.23015$, and the downward translation gives the HR value $-48.71744$ at the same temperature and radii. Between the two curves, the same numerical source favors the upper theory in HR and Einstein in BS. Applying Eq.~\eqref{eq:scheme-source-shift} aligns the curves, including their slopes and kink jumps.

In the $R/L$ clock, the BS potential with source $\xi_{R/L}^{\rm BS}$ and control $U_{R/L}^{\rm BS}=\xi_{R/L}^{\rm BS}\lambda$ is affine in the coupling at fixed radius. Its slope is $3C/R-24\pi C\Theta x-\xi_{R/L}^{\rm BS}$.

On the fixed-$C$ Weyl-squared path, $a_{\rm n}=C$ through first order. The HR vacuum energy $3C/(4R)$ is common to both endpoints, so Eq.~\eqref{eq:fig-scheme-curves} gives coincident curves and triple points. In the continuation of Appendix~\ref{app:weyl-curvature}, the vacuum shift has zero coupling curvature, so the local concavity test applies in both prescriptions.

\section{Weyl-squared geometry and thermodynamic derivatives}
\label{app:weyl}

\subsection{Charged geometry and the first law}

The Weyl-squared dictionary in Sec.~\ref{sec:families} uses the charged solution to first order. In five dimensions its curvature operator is
\begin{equation}
 W_5^2=\mathcal R_{\mu\nu\rho\sigma}^2-\frac43\mathcal R_{\mu\nu}^2
       +\frac16\mathcal R^2 .
\end{equation}
The scalar, Ricci and Riemann squares thus enter in the ratio $(1/6,-4/3,1)$. With these coefficients the charged solution of Cremonini, Liu and Szepietowski \cite{CremoniniLiuSzepietowski2010} takes the form
\begin{gather}
 B_t=B_{t,0}(\rho;\mu,b)+\lambda_{\rm W}B_{t,1}(\rho;\mu,b)+O(\lambda_{\rm W}^2),
 \qquad B_{t,0}=1+\rho^2-\mu/\rho^2+b/\rho^4,\\
 B_{t,1}=-\frac{40b}{\rho^4}+\frac{2\mu^2-20b}{\rho^6}
       +\frac{4\mu b}{\rho^8}+\frac{5b^2}{4\rho^{10}},\qquad
 N=1+\frac{10\lambda_{\rm W}b}{\rho^6}+O(\lambda_{\rm W}^2).
 \label{eq:weyl-solution}
\end{gather}
The gauge field is $\mathcal A_t(\rho)d\widehat t$ in units of $L$. Choose $\mathcal A_t(x)=0$ at the horizon and write $\mathcal A_\infty=\lim_{\rho\to\infty}\mathcal A_t(\rho)$. Integration of $\rho^3\mathcal A_t'/N=P$ gives
\begin{equation}
 \mathcal A_t(\rho)=\mathcal A_\infty-\frac{P}{2\rho^2}-\lambda_{\rm W}\frac{5P^3}{48\rho^8}
 +O(\lambda_{\rm W}^2).
 \label{eq:weyl-gauge}
\end{equation}
The four sectional curvatures in Appendix~\ref{app:stress} give a direct check of the action. Contracting the Riemann, Ricci and scalar terms yields $L^4 W_5^2=2\mathcal W^2$, where $\mathcal W=-X_{t\rho}+X_{t\Omega}+X_{\rho\Omega}-X_{\Omega\Omega}$ in units $L=1$. With $f=f_{\rm W}$, the radial integrand at fixed flux is, up to total derivatives,
\begin{equation*}
 3N\partial_\rho[\rho^2(1-f)+\rho^4]-\frac{NP^2}{2\rho^3}
 +2\lambda_{\rm W}N\rho^3\mathcal W^2 .
\end{equation*}
Varying $f$ and $N$ before expanding in the coupling gives two radial equations. With $\mathcal W_0=-6\mu/\rho^4+15b/\rho^6$ the zeroth-order curvature combination, the first-order coefficients $N_1$ of the lapse and $f_{{\rm W},1}$ of the radial metric function satisfy
\begin{gather*}
 3\rho^2N_1'+180b/\rho^5=0,\\
 -3(\rho^2 f_{{\rm W},1})'+450b^2/\rho^9-432b\mu/\rho^7
 +(480b-24\mu^2)/\rho^5+360b/\rho^3=0 .
\end{gather*}
Both are satisfied by Eq.~\eqref{eq:weyl-solution} with $f_{{\rm W},1}=B_{t,1}-2N_1 B_{t,0}$. The lapse approaches one at infinity, where the mass coefficient defines $\mu$. The correction to $\mathcal A_t'$ in Eq.~\eqref{eq:weyl-gauge} cancels the lapse correction in the flux, giving $\mathcal Q=CP$ per unit sphere volume.

The horizon condition $B_t(x)=0$ fixes $m_1=x^2 B_{t,1}(x;m,b)$. Surface gravity gives $\tau_{\rm cyl}=B_t'(x)/(4\pi N(x))$, with $\mu$ and $b$ held fixed in the derivative. In the Wald formula the Einstein and Weyl-squared curvature derivatives contracted with the binormals give $-2$ and $8\lambda_{\rm W}L^2 W_{\hat0\hat1\hat0\hat1}^{(0)}$, respectively. At zeroth order $\mu=m$, and the normal Weyl component is $L^2W_{\hat0\hat1\hat0\hat1}^{(0)}=\mathcal W_0(x)/2$. The reduced entropy is therefore $4\pi x^3[1-2\lambda_{\rm W}\mathcal W_0(x)]$. The horizon gauge condition in Eq.~\eqref{eq:weyl-gauge} gives the potential coefficients in Eqs.~\eqref{eq:weyl-thermo-zero} and \eqref{eq:weyl-thermo-linear}. Together with the stress-tensor energy, these quantities satisfy
\begin{equation}
 e_x-\tau_{\rm cyl}s_x=O(\lambda_{\rm W}^2),\qquad
 e_P-\tau_{\rm cyl}s_P-\phi=O(\lambda_{\rm W}^2).
\end{equation}
The fixed-charge action also checks the energy. The insertion in Eq.~\eqref{eq:weyl-insertion} is the first-order coefficient of the on-shell reduced free energy at fixed bath temperature. The horizon shift multiplies $e_{0,x}-\tau_0s_{0,x}=0$, so the coefficient is $e_1-\tau_0s_1$ at the reference radius. With the Wald entropy this gives $e_1=w+\tau_0s_1=3m_1+24m$. This agrees with the state-dependent stress-tensor term. The vacuum term is evaluated below.

The grand-canonical calculation of Cremonini, Liu and Szepietowski \cite{CremoniniLiuSzepietowski2010} offers another check. Evaluating their result at the actual horizon radius gives the first-order coefficient of the reduced BS grand potential,
\begin{equation}
 \omega_1=m_1+\frac{40b}{x^2}+\frac{40mb}{x^6}
           -\frac{12m^2}{x^4}-\frac{85b^2}{2x^8}
         =e_1-\tau_0s_1-\tau_1s_0-\phi_1P .
\end{equation}
The Legendre transform adds $\phi_1P$, where $\phi_1$ is the coefficient of $\lambda_{\rm W}$ in $\phi$. Restoring the HR vacuum energy gives the zeroth-order cylinder result. At fixed bath temperature the displacement of the zeroth-order radius cancels $-\tau_1s_0$, giving Eq.~\eqref{eq:weyl-onshell}.

An independent numerical route uses the entropy coordinate $y_s$ defined by $s=4\pi y_s^3$. The first-order off-shell potential is $e_0(y_s)-t_{\rm cyl}s+\lambda_{\rm W}w(y_s,b)$. The two radial stationarity equations and equality of free energies give the check in Sec.~\ref{sec:phase-maps}.

\subsection{Anomaly and cylinder energy checks}

To determine the vacuum energy, we first check the Euler anomaly along the Weyl-squared path. The coefficient $g_{(2)ij}$ of $z_{\rm FG}^2$ in the compactified Einstein metric is fixed by the boundary Schouten tensor $\mathcal S_{ij}^{(4)}$ \cite{deHaro2001},
\begin{equation}
 g_{(2)ij}=-\mathcal S_{ij}^{(4)} .
\end{equation}
In the compactified metric the leading normal Riemann component equals the boundary Schouten tensor. Its contribution to the five-dimensional Schouten tensor cancels the normal Weyl component, and the tangential Weyl components approach the boundary Weyl tensor. Restoring the conformal factor gives $z_{\rm FG}^4 W_4^2/L^4$. The volume element then gives a logarithmic insertion $C\lambda_{\rm W}W_4^2$ in the anomaly convention
\begin{equation}
 \left\langle T^i{}_i\right\rangle=\frac{cW_4^2-aE_4}{16\pi^2}
\end{equation}
where $\langle T^i{}_i\rangle$ is the trace of the renormalized boundary stress tensor and $E_4$ is the boundary Euler density. The insertion shifts $c$ by $\delta c=16\pi^2 C\lambda_{\rm W}$, as in Eq.~\eqref{eq:w-anomalies}.

Both the bulk Weyl tensor and the first variation of the Weyl-squared term vanish on the pure AdS cylinder. The Einstein Brown--York tensor with AdS counterterms therefore determines the vacuum contribution \cite{BalasubramanianKraus1999}. At areal cutoff radius $r$ the zero-type-D vacuum energy density is
\begin{equation}
 8\pi G_5\rho_{\rm cut}
 =-\frac{3}{r}\sqrt{1+r^2/L^2}+\frac3L+\frac{3L}{2r^2}.
\end{equation}
The terms of order $r^0$ and $r^{-2}$ cancel. The remaining coefficient gives the boundary energy predicted by the anomaly,
\begin{equation}
 E_{\rm vac}=\lim_{r\to\infty}\frac{r^4\rho_{\rm cut}}R=\frac{3C}{4R}.
\end{equation}

\subsection{Canonical coexistence near the Weyl-squared critical edge}
\label{app:weyl-canonical}

The Einstein critical point has radius $x_{{\rm c},E}=1/\sqrt{3}$ and metric charge $b_{{\rm c},E}=1/135$. Through first order, write $x_c=x_{{\rm c},E}+\lambda_{\rm W}x_{c,1}$ and $b_c=b_{{\rm c},E}+\lambda_{\rm W}b_{c,1}$. Expanding the two criticality conditions gives Eq.~\eqref{eq:weyl-critical-linear}, with all derivatives evaluated at $(x_{{\rm c},E},b_{{\rm c},E})$.

For the coexistence conditions, write the radii as $x_\pm=\bar x\pm d$, with midpoint $\bar x$ and half-separation $d$. Let $\tau_\pm=\tau_{\rm cyl}(x_\pm)$ and $\bar\tau=(\tau_++\tau_-)/2$. Equation~\eqref{eq:weyl-critical-equal-area} enforces equal temperatures and equal free energies, with $\tau_{\rm cyl}$ and $s_x$ in the integral evaluated at $x=\bar x+du$, where $u$ is the integration variable.

\begin{equation}
 \tau_{0,xb}b_{c,1}+\tau_{1,x}=0,\qquad
 \tau_{0,xxx}x_{c,1}+\tau_{0,xxb}b_{c,1}+\tau_{1,xx}=0.
 \label{eq:weyl-critical-linear}
\end{equation}
\begin{equation}
 \frac{\tau_+-\tau_-}{2d}=0,\qquad
 \frac{1}{d^2}\int_{-1}^{1}(\tau_{\rm cyl}-\bar\tau)s_x\,du=0.
 \label{eq:weyl-critical-equal-area}
\end{equation}

In Eq.~\eqref{eq:weyl-critical-linear}, $\tau_{0,xb}=135/(2\pi)$ and $\tau_{0,xxx}=36/\pi$. The temperature coefficients in Eqs.~\eqref{eq:weyl-thermo-zero} and \eqref{eq:weyl-thermo-linear} give $b_{c,1}=-757/1125$ and $x_{c,1}=-1429\sqrt{3}/300$. The charge follows from $(q_c^{\rm W})^2=3b_c$ and the temperature shift from $\tau_1+\tau_{0,b}b_{c,1}$, giving Eq.~\eqref{eq:weyl-critical-edge}.

Both expressions in Eq.~\eqref{eq:weyl-critical-equal-area} are smooth in $d^2$ as $d$ tends to zero. Their limits are $\tau_{{\rm cyl},x}$ and $2(\tau_{{\rm cyl},x}s_{xx}-\tau_{{\rm cyl},xx}s_x)/3$. Since $s_x>0$, the limiting equations recover the critical conditions. The nonzero derivatives $\tau_{0,xb}$ and $\tau_{0,xxx}$ allow $b$ and $\bar x$ to be solved smoothly in $d^2$ and the coupling near the critical point.

At the local critical edge, denote evaluation at the critical radius and charge by a subscript $c$ on a derivative. The coefficients $(\tau_{{\rm cyl},xb})_c$ and $(\tau_{{\rm cyl},xxx})_c$ are positive at the Einstein point and remain positive at sufficiently small coupling. The expansion of $\tau_{{\rm cyl},x}$ about the critical state then gives two nearby temperature turning points on the subcritical side and none on the supercritical side. Expanding the equal-temperature condition in Eq.~\eqref{eq:weyl-critical-equal-area} at fixed coupling and using $q^2=3b$ gives
\begin{equation*}
 q_c^{\rm W}-q
 =\frac{(\tau_{{\rm cyl},xxx})_c}{4q_c^{\rm W}(\tau_{{\rm cyl},xb})_c}\,d^2
  +O(d^4),\qquad
 s_l-s_s=2(s_x)_c\,d+O(d^3).
\end{equation*}
Thus the entropy gap closes as $\sqrt{q_c^{\rm W}-q}$ at fixed coupling. The regular dependence on $d^2$ makes $q/q_c^{\rm W}$ a regular coordinate there. The chain rule at fixed $q/q_c^{\rm W}$ then gives Eq.~\eqref{eq:weyl-canonical-surface}.

\subsection{Coupling curvature at fixed bath temperature}
\label{app:weyl-curvature}

For the continuation with the Weyl-squared term as the sole correction, the derivative of the reduced canonical free energy $h_{\rm on}$ at fixed cylinder data is the explicit bulk insertion,
\begin{equation}
 \partial_{\lambda_{\rm W}}h_{\rm on}=-\int_{x(\lambda_{\rm W})}^\infty d\rho\,N\rho^3 L^4W_5^2 .
 \label{eq:weyl-full-insertion}
\end{equation}
Here $x(\lambda_{\rm W})$ is the horizon radius of the fixed-temperature saddle. The fixed-charge Maxwell boundary term has no explicit dependence on $\lambda_{\rm W}$, and the round-cylinder counterterms give the same vacuum energy along the path. At large $\rho$ the orthonormal Weyl components fall as $\rho^{-4}$. Their contraction with the leading pure-trace extrinsic curvature vanishes, so the explicit Weyl surface term drops out in the cutoff limit. The bulk integral in Eq.~\eqref{eq:weyl-full-insertion} is convergent. The same insertion follows by varying the coupling in the stationary Noether identity \cite{IyerWald1994}.

The need for only first-order fields can be seen by expanding the fixed-data action $I_0[\varphi]+\lambda_{\rm W}I_1[\varphi]$ about the stationary field configuration $\varphi$ of the two-derivative action $I_0$. Here $I_1$ is the Weyl-squared contribution. Write $\delta I_0=0$, with $\delta$ a field variation, and let $\varphi_1$ be the first-order change of the fields. The linearized stationarity equation is $\delta^2 I_0\cdot\varphi_1+\delta I_1=0$. The second on-shell coupling derivative is therefore $\delta I_1\cdot\varphi_1$. This argument parallels the next-to-next-to-leading-order calculation by Ma, Pang and L\"u for asymptotically flat black holes \cite{MaPangLu2023}. With fixed AdS cylinder data, the first-order variation includes the displacement of the smooth horizon.

For the metric in Eq.~\eqref{eq:weyl-metric}, the sectional curvatures give $L^4W_5^2=2\mathcal W^2$ with $\rho$ derivatives denoted by primes,
\begin{equation}
 \mathcal W=\frac{B_t''/2-B_t'/\rho+B_t/\rho^2}{N^2}-\frac1{\rho^2}
   +\frac{N'}{N^3}\left(\frac{B_t}{\rho}-\frac{B_t'}2\right).
\end{equation}
Use Eq.~\eqref{eq:weyl-solution} at fixed $x,b$. Let $\mathcal W_0$ and $\mathcal W_1$ be the zeroth- and first-order coefficients of $\mathcal W$, and let $N_1$ be the first-order lapse coefficient. The variation of the insertion at fixed horizon radius is
\begin{equation}
 K_{\rm ins}=-\int_x^\infty d\rho\,\rho^3(2N_1\mathcal W_0^2+4\mathcal W_0\mathcal W_1),
 \qquad N_1=10b/\rho^6 .
 \label{eq:weyl-insertion-linear}
\end{equation}
Evaluating the integral at fixed radius and charge gives, with $y=x^2$ and $u=b/x^4$,
\begin{equation*}
 K_{\rm ins}=\frac{12}{7y}\left[56(1+y)^3
 -28u(2y^2+9y+7)+35u^2(3y+8)-158u^3\right].
\end{equation*}
On a thermal saddle the fixed-temperature derivative equates the insertion to $R\Xi_{\rm cyl}^{\rm W}/C$. Choosing the bath temperature equal to the state temperature at a prescribed radius gives the expansion in Sec.~\ref{sec:families}.
Along the fixed-temperature branch the horizon displacement adds $w_x x_1$ to the derivative of the insertion. The relaxed and fixed-radius curvatures are then
\begin{equation}
 K_t=K_{\rm ins}-w_x\frac{\tau_1}{\tau_0'},\qquad K_x=K_{\rm ins}+\tau_1s_1 .
\end{equation}
The fixed-radius identity follows by differentiating $e_\lambda-\tau_{\rm cyl}s_\lambda$ at fixed radius. With primes now denoting $x$ derivatives, the radial first law gives $w_x=\tau_1s_0'-\tau_0's_1$. Substitution yields Eq.~\eqref{eq:radial-relaxation}.

In terms of $y=x^2$ and $u=b/x^4$, the numerator in Eq.~\eqref{eq:weyl-curvature} is
\begin{align}
 \mathcal P(y,u)={}&112y^2(y-u/2)^2+448y^3+(504u^2-728u+672)y^2 \notag\\
 &+(198u^3+910u^2-1120u+448)y+p_4(u),\\
 p_4(u)={}&117u^4-554u^3+868u^2-504u+112 .
\end{align}
The quadratic in the $y^2$ coefficient and the quadratic part of the $y$ coefficient have negative discriminants and positive leading coefficients. The Sturm chain of $p_4$ has signs $(+,-,+,+,+)$ at zero and $(+,+,+,-,+)$ at positive infinity. The variation counts agree. Since $p_4(0)=112$, the quartic is positive on the nonnegative real axis, which gives the sign of the numerator in Eq.~\eqref{eq:weyl-curvature}. The fixed-radius curvature is
\begin{equation}
 K_x=\frac{12u}{7y}\left[73u^2-49uy-84u+28(1+y)^2\right].
\end{equation}

An independent interaction at the same perturbative order illustrates the dependence on the chosen continuation. Add $\kappa\lambda_{\rm W}^2L^4(F^2)^3$ to Eq.~\eqref{eq:actions}, where $\kappa$ is an independent dimensionless coefficient. The direct insertion on the Einstein saddle shifts $K_t$ by $8\kappa P^6/(7x^{14})$. To obtain it, insert $F^2=-2P^2/(L^2\rho^6)$ for the analytically continued electric field and integrate the radial power. The extra flux boundary term vanishes at infinity. The term enters the second-order curvature while the linear thermodynamic coefficients retain the values in Sec.~\ref{sec:families}.

\bibliographystyle{JHEP}
\bibliography{references}
\end{document}